\documentclass[twocolumn,aps,prl,superscriptaddress,tightenlines]{revtex4}
\usepackage[latin9]{inputenc}
\usepackage{amsmath}
\usepackage{amssymb}
\usepackage{graphicx}
\usepackage{esint}

\makeatletter
\@ifundefined{textcolor}{}
{%
 \definecolor{BLACK}{gray}{0}
 \definecolor{WHITE}{gray}{1}
 \definecolor{RED}{rgb}{1,0,0}
 \definecolor{GREEN}{rgb}{0,1,0}
 \definecolor{BLUE}{rgb}{0,0,1}
 \definecolor{CYAN}{cmyk}{1,0,0,0}
 \definecolor{MAGENTA}{cmyk}{0,1,0,0}
 \definecolor{YELLOW}{cmyk}{0,0,1,0}
}

\usepackage{mathrsfs}\usepackage{bbm}\usepackage{bbold}\usepackage{subfigure}

\newcommand{\bk}{{\bf k}}

\newcommand{\dif}{\mathrm{d}}

\makeatother

\begin{document}

\title{Manipulation of gap nodes by uniaxial strain in iron-based superconductors}

\author{Jian Kang}

\affiliation{School of Physics and Astronomy, University of Minnesota, Minneapolis,
MN 55455, USA}

\author{Alexander F. Kemper}

\affiliation{Lawrence Berkeley National Lab, 1 Cyclotron Road, Berkeley, California
94720, USA}

\author{Rafael M. Fernandes}

\affiliation{School of Physics and Astronomy, University of Minnesota, Minneapolis,
MN 55455, USA}
\begin{abstract}
In the iron pnictides and chalcogenides, multiple orbitals participate
in the superconducting state, enabling different gap structures to
be realized in distinct materials. Here we argue that the spectral
weights of these orbitals can in principle be controlled by a tetragonal
symmetry-breaking uniaxial strain, due to the enhanced nematic susceptibility
of many iron-based superconductors. By investigating multi-orbital
microscopic models in the presence of orbital order, we show that
not only $T_{c}$ can be enhanced, but pairs of accidental gap nodes
can be annihilated and created in the Fermi surface by an increasing
external strain. We explain our results as a mixture of nearly-degenerate
superconducting states promoted by strain, and show that the annihilation
and creation of nodes can be detected experimentally via anisotropic
penetration depth measurements. Our results provide a promising framework
to externally control the superconducting properties of iron-based
materials.
\end{abstract}
\maketitle
A distinguishing feature of iron pnictides and chalcogenides \cite{LaOFeAs,BaFe2As2}
is the non-universality of their gap structures. Experimentally, nodeless
superconducting (SC) gaps are observed in optimally doped $\mathrm{Ba_{1-x}K_{x}Fe_{2}As_{2}}$
\cite{DingARPES122K,Shimojima_Kdoped} and $\mathrm{Ba(Fe_{1-x}Co_{x})_{2}As_{2}}$
\cite{Prozorov122CoGap}, as well as in the undoped material $\mathrm{LiFeAs}$
\cite{BorisenkoLi111Gap,AllanLiFeAs}, while gap nodes are reported
in optimally doped $\mathrm{BaFe_{2}(As_{1-x}P_{x})_{2}}$ \cite{Matsuda122PNode,FengPnode,Pdoped_nodes},
$\mathrm{Ba(Fe_{1-x}Ru_{x})_{2}As_{2}}$ \cite{Ru_doped_nodes}, and
in the parent compounds $\mathrm{FeSe}$ \cite{Xue11Node} and $\mathrm{KFe_{2}As_{2}}$
\cite{MatsudaK122Node,ShinK122Node,TailleferK122Pairing}. This diversity
of behaviors opens up the interesting possibility of manipulating
the superconducting ground state by tuning the appropriate external
parameters. While this can be achieved empirically by mixing different
types of doping, such as $\mathrm{Ba(Fe_{1-x}Co_{x})_{2}(As_{1-y}P_{y})_{2}}$
\cite{Johrendta122CoP}, control of the SC state requires understanding
the mechanisms responsible for this non-universality of the gap structure.

\begin{figure}[htbp]
\centering \subfigure[\label{Fig:OOEffect:FS}]{\includegraphics[width=0.6\columnwidth]{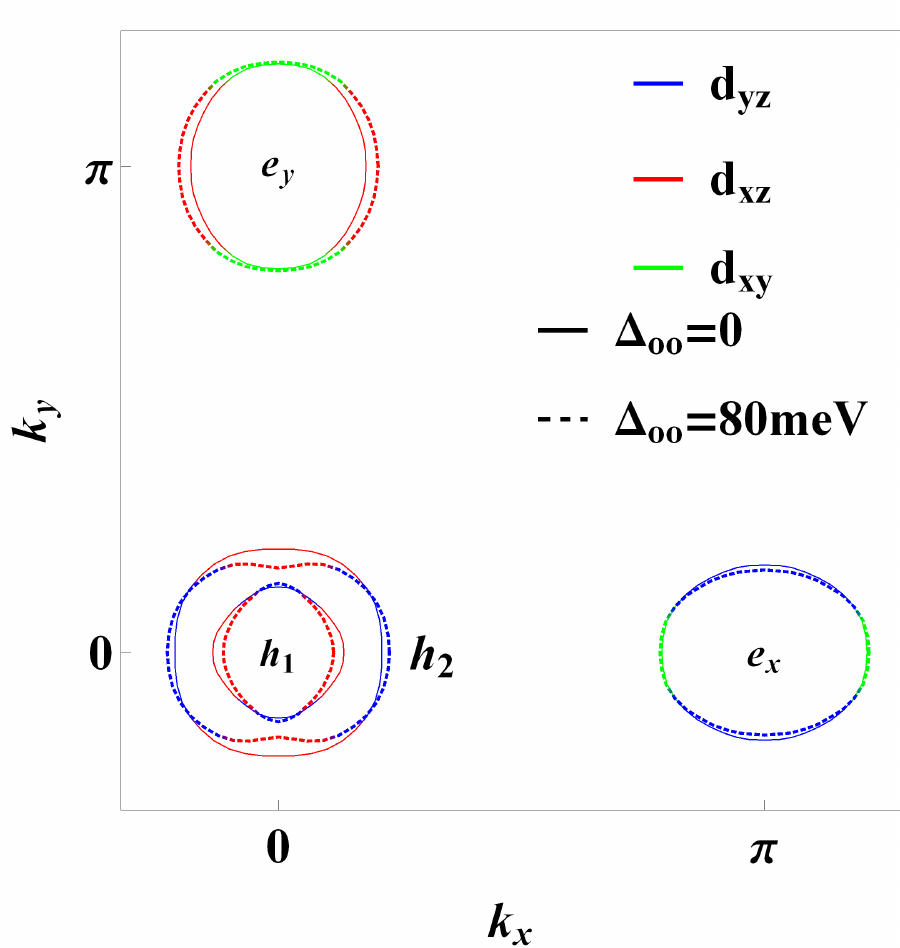}}
\\
 \subfigure[\label{Fig:OOEffect:Tc}]{\includegraphics[width=0.5\columnwidth]{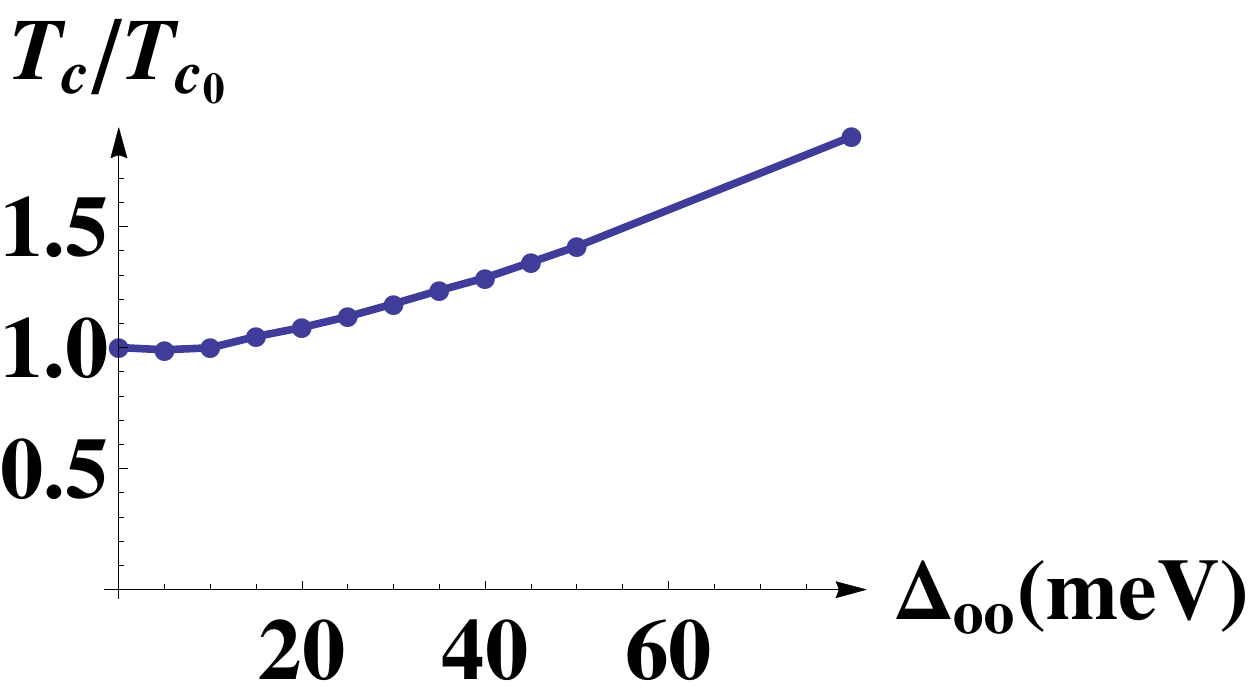}}
\caption{(a) Change of the FS by orbital order. The solid (dashed) lines describe
the FS for $\Delta_{oo}=0$ ($\Delta_{oo}=80$ meV). Different colors
represent the dominant $3d$ orbital component on the FS. (b) Enhancement
of $T_{c}\propto\mathrm{e}^{-1/\lambda_{\alpha}}$ (in units of its
value in the tetragonal phase $T_{c_{0}}$) as function of the orbital
order parameter $\Delta_{oo}$. \label{Fig:OOEffect}}
\end{figure}

Theoretically, spin fluctuations have been widely proposed to cause
pairing in iron pnictides and chalcogenides \cite{magnetic}. In this
model, the non-universal behavior of the gap structure stems from
the multi-orbital character of these materials that arises due to
the $3d^{6}$ configuration of Fe \cite{reviews_pairing}. In fact,
first-principle calculations \cite{GraserSDDeg} and ARPES experiments
\cite{DingARPESOrbital,Shen122CoOO} reveal that the disconnected
pockets that form the Fermi surface of most pnictides contain significant
spectral weight from the $d_{xz}$, $d_{yz}$, and $d_{xy}$ orbitals
(see Fig. \ref{Fig:OOEffect}(a)). While a sign-changing $s^{+-}$
state is favored by pairing within the $d_{xz}$ and $d_{yz}$ orbitals
of different pockets, a $d$-wave state is preferred by the $d_{xy}$
orbitals. Thus, not only the leading SC instability, but also the
presence or absence of nodes, depends on the orbital content of the
Fermi pockets \cite{Kuroki09,GraserSDDeg,Chubukov09}. Such a near-degeneracy
between different SC ground states, supported by theoretical \cite{CWu09,KemperNJP,Graser10,Ikeda10,Wang10,Maiti11,Thomale11,Fernandes13,Kotliar13}
and experimental results \cite{raman_mode1,raman_mode2}, is a distinguishing
feature of the iron-based materials, since in most superconductors
one SC state usually has a much lower energy than all the other ones.

Therefore, in this framework, the properties of the SC state of the
iron pnictides could be manipulated if the orbital content of their
Fermi surface could be tuned. In this paper, we propose that this
can be achieved via application of uniaxial strain $\partial_{i}u_{i}$,
where $\mathbf{u}$ denotes the displacement vector. Experimentally,
many optimally-doped iron-based superconductors display a large nematic
susceptibility $\chi_{\mathrm{nem}}$ \cite{FisherXnemDivergent,Kuo13,shear_modulus,Yoshizawa12,Matsuda12,Meingast122Xnem,GallaisRamanNem,Jigang14,PasupathyNa111NemAFM},
implying that even a small uniaxial strain \cite{Fisher10,Tanatar10,Degiorgi12,Dhital12}
can trigger a nematic state with sizable anisotropies in the lattice
and, more interestingly, in the magnetic and orbital degrees of freedom
\cite{RMFRevNem}. While previous works investigated how superconductivity
is affected by the nematic-induced anisotropy in the magnetic spectrum
\cite{RMFSDMix}, little is known about the impact of the induced
anisotropy in the electronic spectrum. Indeed, in the nematic state,
the onsite energies of the $d_{xz}$ and $d_{yz}$ orbitals become
unequal, $\Delta_{oo}=\left\langle \varepsilon_{xz}\right\rangle -\left\langle \varepsilon_{yz}\right\rangle \neq0$
\cite{Shen122CoOO,Yi12,Zhang12}, affecting the orbital content of
the Fermi surface \cite{w_ku10,Devereaux10,Phillips12,Kontani12,Dagotto13,Fernandes12}.
Due to the large nematic susceptibility $\chi_{\mathrm{nem}}$, a
sizeable orbital splitting $\Delta_{oo}\sim50$ meV can be triggered
even by a small uniaxial strain of the order of $10$ MPa \cite{Shen122CoOO,Yi12}.
Therefore, because of the sensitivity of the pairing state to the
orbital content of the Fermi surface, strain can be a viable tuning
parameter to manipulate the SC ground state.

Here, using a multi-orbital microscopic model, we show that by changing
the $d_{xz}/d_{yz}$ orbital splitting via uniaxial strain $\Delta_{oo}\propto\left(\partial_{x}u_{x}-\partial_{y}u_{y}\right)\chi_{\mathrm{nem}}$,
gap nodes can be created on a nodeless SC state or manipulated in
a nodal SC state. Focusing on the latter case, we find that, as $\Delta_{oo}$
is enhanced, while pairs of accidental nodes are annihilated in the
electron-like Fermi pockets, merging along the direction of the applied
strain, pairs of nodes are created in the hole-like Fermi pockets,
emerging along both $x$ and $y$ directions. Interestingly, an enhancement
of $T_{c}$ accompanies the motion of the nodes. We argue that these
behaviors are consistent with a sizable mixture between $s^{+-}$
and $d$-wave states promoted by orbital order, which is only meaningful
because of the near-degeneracy between the two states. We also show
that the annihilation and the creation of nodes can be detected experimentally
by sharp features that arise in the penetration depth.

Our starting point is the five-orbital Hubbard model with all possible
on-site interactions \cite{KemperNJP,reviews_pairing}:

\begin{align}
H= & \sum_{ii',ll'}\sum_{\sigma}t_{il,i'l'}c_{il\sigma}^{\dagger}c_{i'l'\sigma}+U\sum_{i,l}n_{il\uparrow}n_{il\downarrow}\label{Eqn:Hamiltonian}\\
 & +U'\sum_{i,l<l'}n_{il}n_{il'}+J\sum_{i,l<l'}\sum_{\sigma,\sigma'}c_{il\sigma}^{\dag}c_{il'\sigma'}^{\dag}c_{il\sigma'}c_{il'\sigma}\nonumber \\
 & +J'\sum_{il'\neq l}c_{il\uparrow}^{\dag}c_{il\downarrow}^{\dag}c_{il'\downarrow}c_{il'\uparrow}+\frac{\Delta_{oo}}{2}\sum_{i}(n_{i,yz}-n_{i,xz})\ .\nonumber
\end{align}

Here, $c_{il\sigma}^{\dagger}$ creates an electron at orbital $l$
and site $i$, and $n_{il\sigma}=c_{il\sigma}^{\dagger}c_{il\sigma}^{\phantom{\dagger}}$.
The first term describes the band structure, with tight-binding parameters
fitted to first-principle calculations (see Ref. \cite{KemperNJP}
for the hopping parameters). $U=1.3$ eV is the intra-orbital repulsion,
$J=0.2$ eV is the Hund's rule coupling, $J'=J$ is the pair-hopping
coupling, and $U'=U-2J$ is the inter-orbital repulsion. The orbital
order parameter $\Delta_{oo}$ gives the splitting between the $d_{xz}$
and $d_{yz}$ onsite energies. In the following, we consider that
it is generated by uniaxial strain applied parallel to the $x$ direction,
$\partial_{x}u_{x}>\partial_{y}u_{y}$, implying $\Delta_{oo}>0$
\cite{Shen122CoOO}. Although strain also affects the onsite energies
and hopping parameters of other orbitals, their impact on the electronic
structure is small compared to the contribution arising from the $d_{xz}$-$d_{yz}$
orbital splitting \cite{Shen122CoOO}.

\begin{figure}[htbp]
\centering \subfigure[\label{Fig:SCGapOO:Gap}]{\includegraphics[width=0.95\columnwidth]{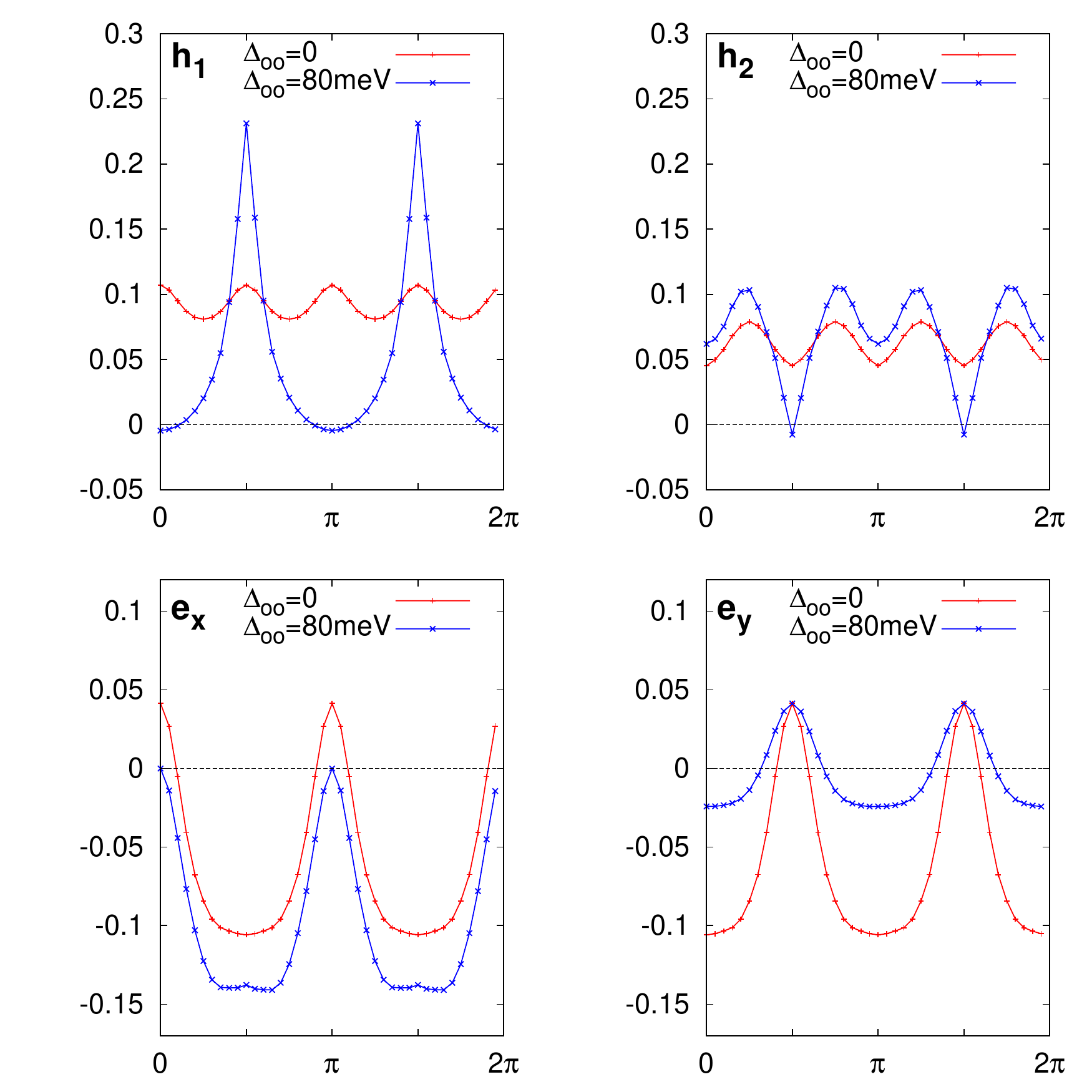}}

\subfigure[\label{Fig:SCGapOO:FSNode}]{\includegraphics[width=0.55\columnwidth]{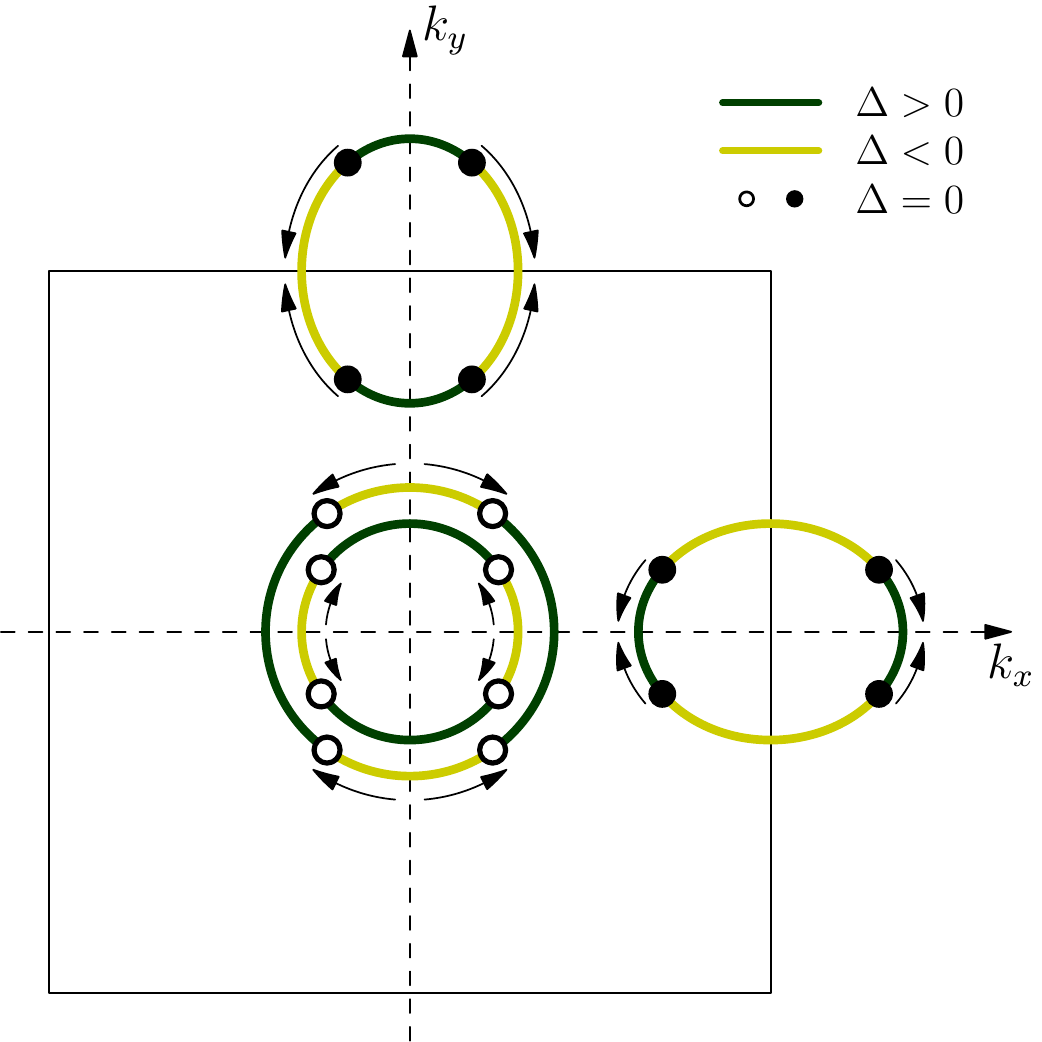}}
\caption{Evolution of gap structure at $T_{c}$ for increasing orbital order
parameter $\Delta_{oo}$. The four panels in (a) show the angular
dependencies of the gaps of each pocket in the presence (blue lines)
and in the absence (red lines) of orbital order. The angles are measured
relatively to the $x$ axis. (b) Schematic illustration of the motion
of nodes as function of increasing $\Delta_{oo}$. While nodes are
created in the hole pockets, emerging along both $x$ and $y$ directions
(white dots), the electron pockets' nodes are annihilated, merging
along the $x$ direction (black dots). Green (yellow) lines denote
positive (negative) gap values. \label{Fig:SCGapOO}}
\end{figure}

The Fermi surface (FS) of this model for an occupation number $n\approx6$
is displayed in Fig~\ref{Fig:OOEffect}(a). In the tetragonal phase
($\Delta_{oo}=0$) the FS is composed of two $C_{4}$-symmetric central
hole pockets ($h_{1}$, $h_{2}$) and two $C_{2}$-symmetric electron
pockets ($e_{x}$, $e_{y}$) centered at $\left(\pi,0\right)$ and
$\left(0,\pi\right)$. While $h_{1}$ and $h_{2}$ have only $d_{xz}$/$d_{yz}$
orbital character, $e_{x}$ has $d_{yz}/d_{xy}$ character and $e_{y}$,
$d_{xz}/d_{xy}$ \cite{GraserSDDeg,KemperNJP}. Consequently, for
a non-zero orbital order parameter $\Delta_{oo}$, the sizes of the
two electron pockets become slightly different, and the two hole pockets
are distorted into $C_{2}$-symmetric shapes \cite{Fernandes12,Lv11}.
To investigate the effect of orbital order on SC, we solve numerically
the linearized spin-fluctuation RPA gap equations \cite{KemperNJP,reviews_pairing}
\begin{equation}
-\sum_{j}\oint_{C_{j}}\frac{\mathrm{d}\bk'}{2\pi}\frac{1}{v_{F}(\bk')}\Gamma^{ij}(\bk,\bk')g_{\alpha j}(\bk')=\lambda_{\alpha}g_{\alpha i}(\bk)\ ,\label{Eqn::MultiSCEqn}
\end{equation}
 where $v_{F}(\bk)$ is the Fermi velocity, $C_{j}$ denotes one of
the four Fermi pockets, and $\Gamma^{ij}(\bk,\bk')$ is the effective
pairing interaction which scatters a Cooper pair $(\bk,-\bk)$ on
the FS $C_{i}$ to $(\bk',-\bk')$ on the FS $C_{j}$. The structure
factor of the SC gap at pocket $C_{i}$ is given by $g_{\alpha i}\left(\mathbf{k}\right)$,
and the largest eigenvalue $\lambda_{\alpha}$ gives the leading pairing
instability, with $T_{c}\propto\mathrm{e}^{-1/\lambda_{\alpha}}$.
For $\Delta_{oo}=0$, as shown previously \cite{GraserSDDeg,KemperNJP},
the leading pairing instability is the $s^{+-}$ state with accidental
nodes on the two electron pockets (see red lines in Fig. \ref{Fig:SCGapOO}),
followed closely by a $d$-wave state with symmetry-constrained nodes
along the diagonals of the Brillouin zone.

Solving the linearized gap equations in the presence of a non-zero
orbital order $\Delta_{oo}\neq0$ for a fixed occupation number, we
find a steady enhancement of $T_{c}\propto\mathrm{e}^{-1/\lambda_{\alpha}}$
for increasing $\Delta_{oo}$, as shown in Fig. \ref{Fig:OOEffect}(b).
The gap structure is also strongly affected by orbital order: in Fig.
\ref{Fig:SCGapOO}(a), we contrast the angular dependence of the gap
around each of the Fermi pockets in the tetragonal phase ($\Delta_{oo}=0$)
and in the presence of orbital order ($\Delta_{oo}=80$ meV, motivated
by the experimentally measured values \cite{Shen122CoOO,Yi12}). Clearly,
while in the $h_{1}$ and $h_{2}$ hole pockets the gaps become more
anisotropic, in the $e_{x}$ and $e_{y}$ electron pockets they become
more isotropic. Consequently, for increasing $\Delta_{oo}$, pairs
of accidental gap nodes tend to be created in the hole pockets, emerging
parallel (perpendicular) to the strain direction in $h_{1}$ ($h_{2}$),
while the pairs of accidental nodes initially present in the electron
pockets tend to be annihilated, merging along the strain direction
for both $e_{x}$ and $e_{y}$. An schematic illustration of this
nodal behavior is shown in Fig. \ref{Fig:SCGapOO}(b).

To show that these results are general and do not depend on details
of the tight-binding model, we interpret them as a consequence of
the mixing between the $s^{+-}$ and $d$-wave gap functions of the
tetragonal state promoted by orbital order \cite{RMFSDMix}:
\begin{equation}
\Delta\approx\Delta_{s}+\gamma\Delta_{d}\ ,\label{Eqn::SDGap}
\end{equation}
 where $\gamma\propto\Delta_{oo}$ is the mixing parameter, which
is sensitive to the applied strain and to the nematic susceptibility,
since $\Delta_{\mathrm{oo}}\propto\left(\partial_{x}u_{x}-\partial_{y}u_{y}\right)\chi_{\mathrm{nem}}$.
Of course, by symmetry, the gap function of any orthorhombic system
is a mixture of $s$-wave and $d$-wave states. What makes the pnictides'
case interesting, and somehow unique, is the near-degeneracy between
the $s^{+-}$ and $d$-wave states, which enforces the mixing parameter
$\gamma$ to be sizable. This is to be contrasted to the case of orthorhombic
cuprates, where the $s$-wave component arising from the orthorhombic
symmetry is, for most purposes, irrelevant.

One of the consequences of the near-degeneracy between the competing
$s^{+-}$ and $d$-wave states is the suppression of the value of
$T_{c}$ in the tetragonal phase \cite{Fernandes13,RMFSDMix}. The
mixing between the two states promoted by orbital order, Eq. (\ref{Eqn::SDGap}),
lifts this degeneracy, which leads to an effective enhancement of
$T_{c}$ \cite{Fernandes13,RMFSDMix}, as found in Fig. \ref{Fig:OOEffect}(b).
Interestingly, our RPA calculation suggests an enhancement that can
be as large as $50\%$ for realistic values of $\Delta_{oo}$. Furthermore,
Eq. (\ref{Eqn::SDGap}) also explains qualitatively the motion of
the gap nodes displayed in Fig. \ref{Fig:SCGapOO} as the natural
evolution from a nodal $s^{+-}$ to a $d$-wave state. To illustrate
this, consider simple harmonic expressions for the gaps in the tetragonal
nodal $s^{+-}$ state, $\Delta_{h_{1/2}}^{(s)}=\Delta_{0}$ and $\Delta_{e_{x/y}}^{(s)}=-\Delta_{1}\pm\Delta_{2}\cos2\phi_{x/y}$.
Here, $\Delta_{j}>0$ and $\Delta_{2}>\Delta_{1}$; $\phi_{x/y}$
and $\theta$ denote the polar angles along the electron and hole
pockets, respectively, measured with respect to the $ $$x$ axis.
The presence of orbital order gives rise to additional $d$-wave components
$\Delta_{h_{1/2}}^{(d)}=\mp\tilde{\Delta}_{0}\cos2\theta$ and $ $$\Delta_{e_{x/y}}^{(d)}=\mp\tilde{\Delta}_{1}$
(with $\tilde{\Delta}_{j}>0$) in the gap functions. Thus, according
to Eq. (\ref{Eqn::SDGap}), as the mixing parameter $\gamma$ increases,
the pairs of nodes initially present in the electron pockets eventually
merge along the $x$ direction and disappear, while pairs of nodes
appear in the hole pocket $h_{1}$ ($h_{2}$) along the $x$ ($y$)
direction.

Whereas the above argument successfully accounts for the RPA calculations,
a natural limitation of the latter is their validity only near $T_{c}$.
An important question is whether the gap structure obtained at $T_{c}$
is robust down to $T=0$. To address this question, we developed an
effective three-orbital model, inspired by the RPA analysis at $T_{c}$.
In this model, only the $d_{xz}$, $d_{yz}$, and $d_{xy}$ orbitals
are included, since their spectral weights dominate the FS (see Fig.
\ref{Fig:OOEffect}). Furthermore, since the RPA-derived pairing interaction
$\Gamma^{ij}(\bk,\bk')$ is dominated by $i=h_{1}$ and $j=e_{x/y}$
\cite{KemperNJP}, we consider these three pockets only. We also focus
on the anisotropies introduced by the orbital contents of the FS pockets,
rather than on their shapes. Consequently, we assume circular pockets
and expand their wave-functions in terms of harmonic functions of
their polar angles $\theta$ and $\phi_{x/y}$ (see also Refs. \cite{Maiti11,Vafek13}):
\begin{align}
 & \left|h\right\rangle =\cos\theta\left(1+\beta_{h}\sin^{2}\theta\right)\left|d_{xz}\right\rangle +\sin\theta\left(1-\beta_{h}\cos^{2}\theta\right)\left|d_{yz}\right\rangle \nonumber \\
 & \left|e_{x/y}\right\rangle =\alpha\left|d_{xy}\right\rangle +\left(1\pm\beta_{e}\right)\left\{ \begin{array}{c}
\sin\phi_{x}\\
\cos\phi_{y}
\end{array}\right\} \left|d_{yz/xz}\right\rangle \label{EQ_effective_model}
\end{align}

Here, the parameter $\alpha=0.52$ is obtained by fitting the angular
dependence of $\left\langle \left.d_{a}\right|e_{x/y}\right\rangle $
to the corresponding tight-binding matrix elements for $\Delta_{oo}=0$
and the parameters $\beta_{h}$, $\beta_{e}=2\beta_{h}$ describe
how the orbital weight around each pocket is changed by orbital order
(see the supplementary material).

\begin{figure}[htbp]
\begin{centering}
\includegraphics[width=1.4in]{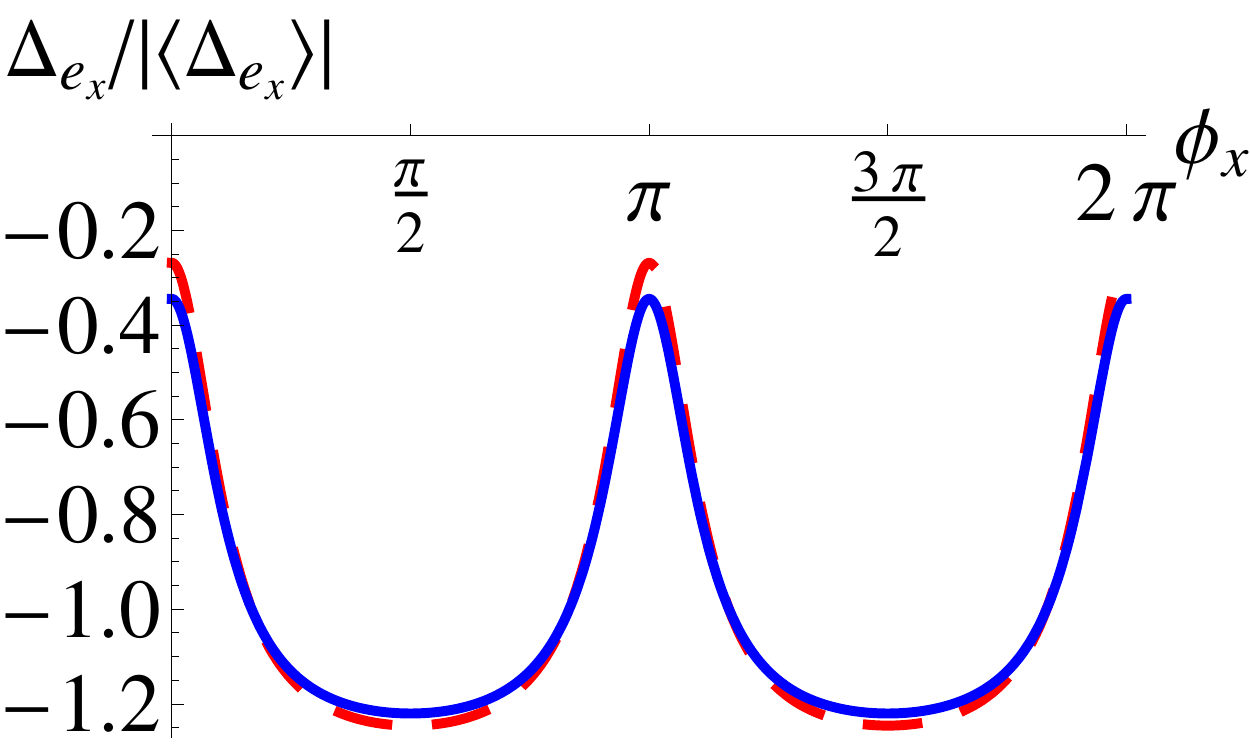}\hfill{}\includegraphics[width=1.4in]{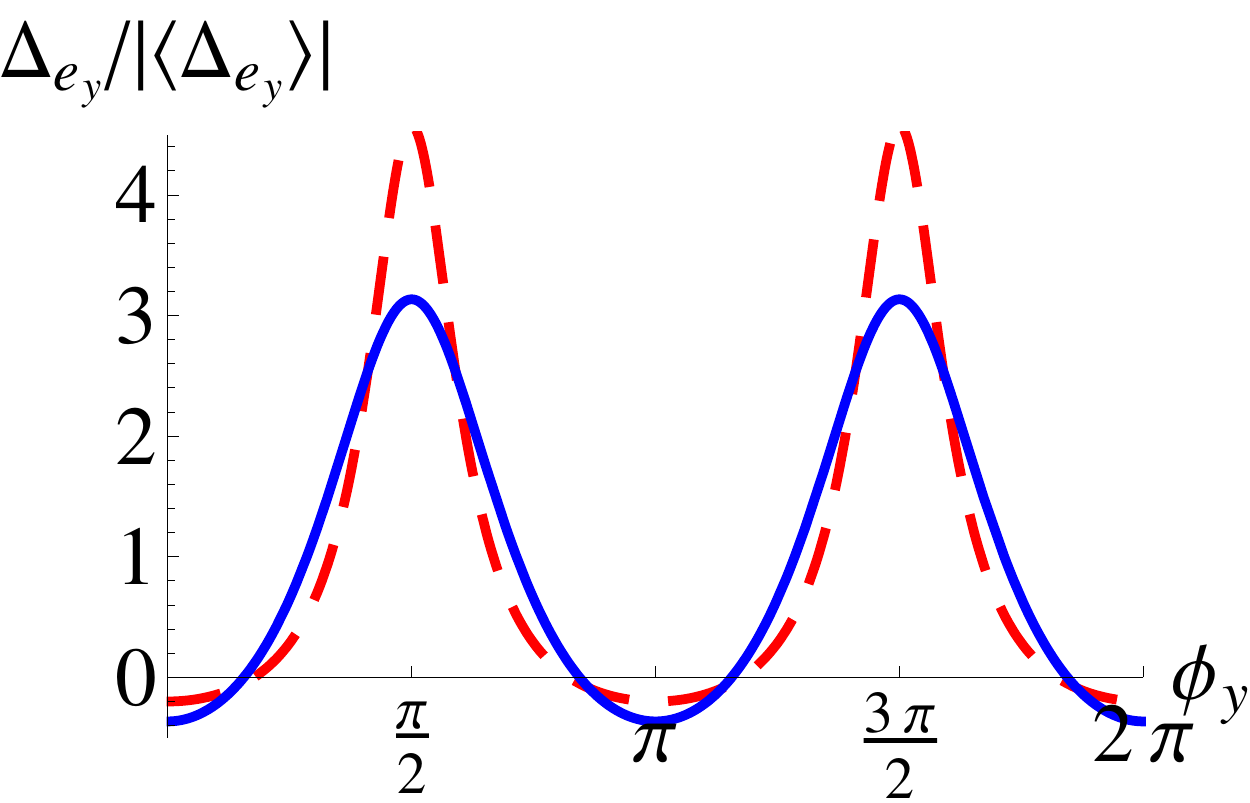}
\par\end{centering}

\smallskip{}

\centering{}\includegraphics[width=1.4in]{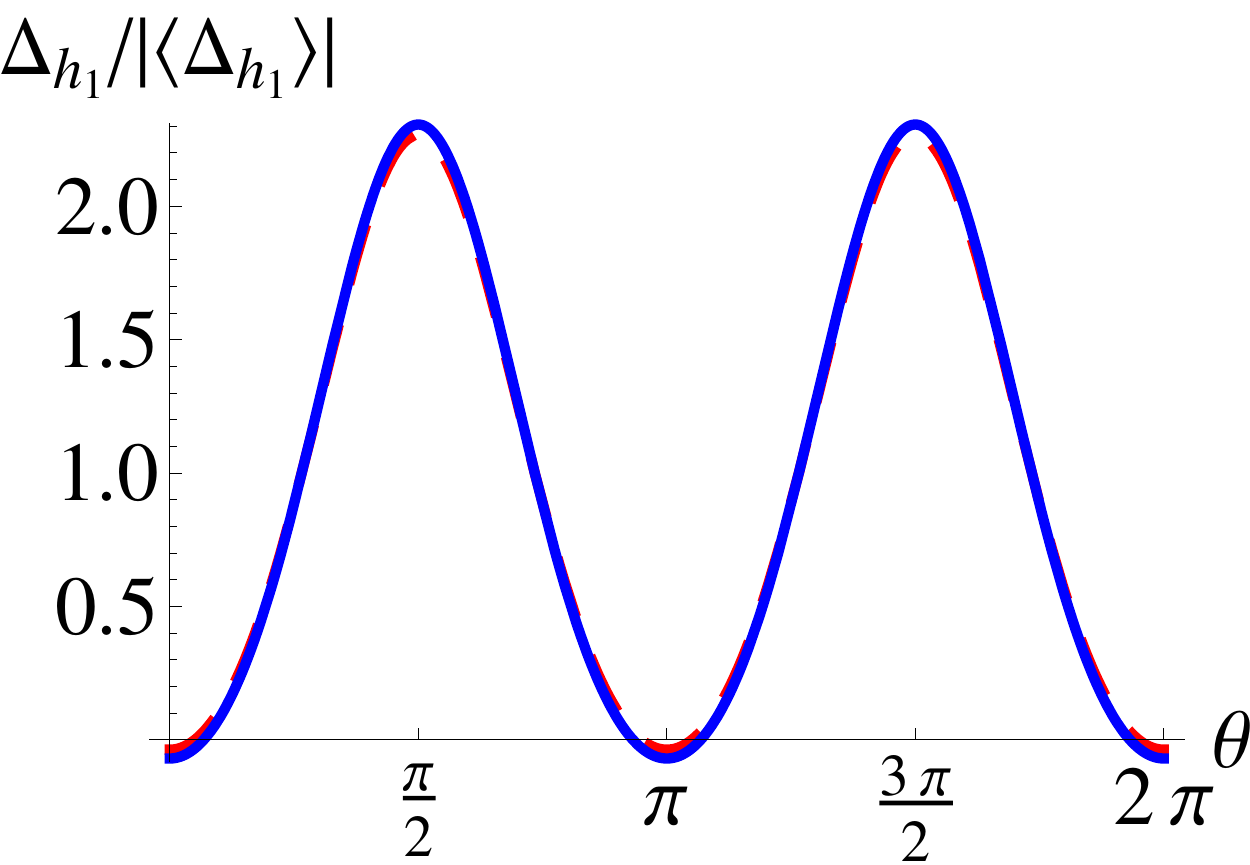} \caption{Angular dependence of the gaps on the hole ($h_{1}$) and electron
($e_{x/y}$) pockets at $T_{c}$ (dashed red line) and at $T=0$ (solid
blue line) within our effective three-band model. The gaps are normalized
by their absolute average values. The parameters used are: $v=0.30$,
$w=0.25$, $\beta_{e}=2\beta_{h}=0.40$. \label{Fig:Model} }
\end{figure}

For the pairing interaction, motivated once again by the RPA results
\cite{KemperNJP}, we consider only the intra-orbital pairing interactions
$v$ and $w$ connecting, respectively, the $d_{xz/yz}$ and $d_{xy}$
orbitals from different pockets. It is then straightforward to write
down the BCS-like gap equations and solve them at any temperature
(details in the supplementary material). For $\Delta_{oo}=0$ and
at $T=T_{c}$, an $s^{+-}$ state with accidental nodes on the electron
pockets is found for $v\gtrsim w$, whereas a $d$-wave state is found
for $v\lesssim w$. To capture the near degeneracy between these two
states, we focus on the regime $v\approx w$ -- similar to the RPA
case -- and solve the gap equations for $T=0$ and $\Delta_{oo}\neq0$.
Fig.~\ref{Fig:Model} contrasts the angular dependency of the gaps
at $T=T_{c}$ and $T=0$, evidencing the robustness of the gap structure
obtained at $T_{c}$. Furthermore, we also find that $T_{c}$ increases
with increasing $\Delta_{oo}$, demonstrating that our effective model
captures the RPA-derived results. An important issue is whether the
$s$-$d$ mixing parameter $\gamma$ in Eq.~\ref{Eqn::SDGap} is
always real, or whether imaginary solutions may arise, resulting in
time-reversal symmetry-breaking states \cite{Valentin,Maiti}. Phenomenologically,
the coupling between orbital order and SC gives rise to the quadratic
free energy term $F\propto\Delta_{\mathrm{oo}}\left|\Delta_{s}\right|\left|\Delta_{d}\right|\cos\theta$,
where $\theta$ is the relative phase between the SC order parameters,
which is minimized by $\theta=0$ or $\theta=\pi$, i.e. by a real
admixture of the $s^{+-}$ and d-wave states. In the supplementary
material, we show explicitly that in our microscopic model $\Delta_{\mathrm{oo}}\neq0$
favors a real $\gamma$ for all temperatures.

Having established that accidental nodes can be manipulated by uniaxial
strain via the induced orbital order at all temperatures, we now discuss
their experimental manifestations. As shown in Fig. \ref{Fig:SCGapOO},
when pairs of nodes are annihilated (created) in the electron (hole)
pockets, they merge into (emerge from) a single node with quadratic
quasi-particle dispersion. These quadratic nodes give rise to a density
of states $N\left(\omega\right)$ that scales as $\sqrt{\omega}$
at low energies \cite{quadratic_nodes1,quadratic_nodes2,quadratic_nodes3}
-- in contrast to the usual behavior for linear nodes $N\left(\omega\right)\sim\omega$
-- strongly affecting the low-temperature behavior of thermodynamic
quantities in the superconducting state.

\begin{figure}[htbp]
\centering{}\includegraphics[width=0.57\columnwidth]{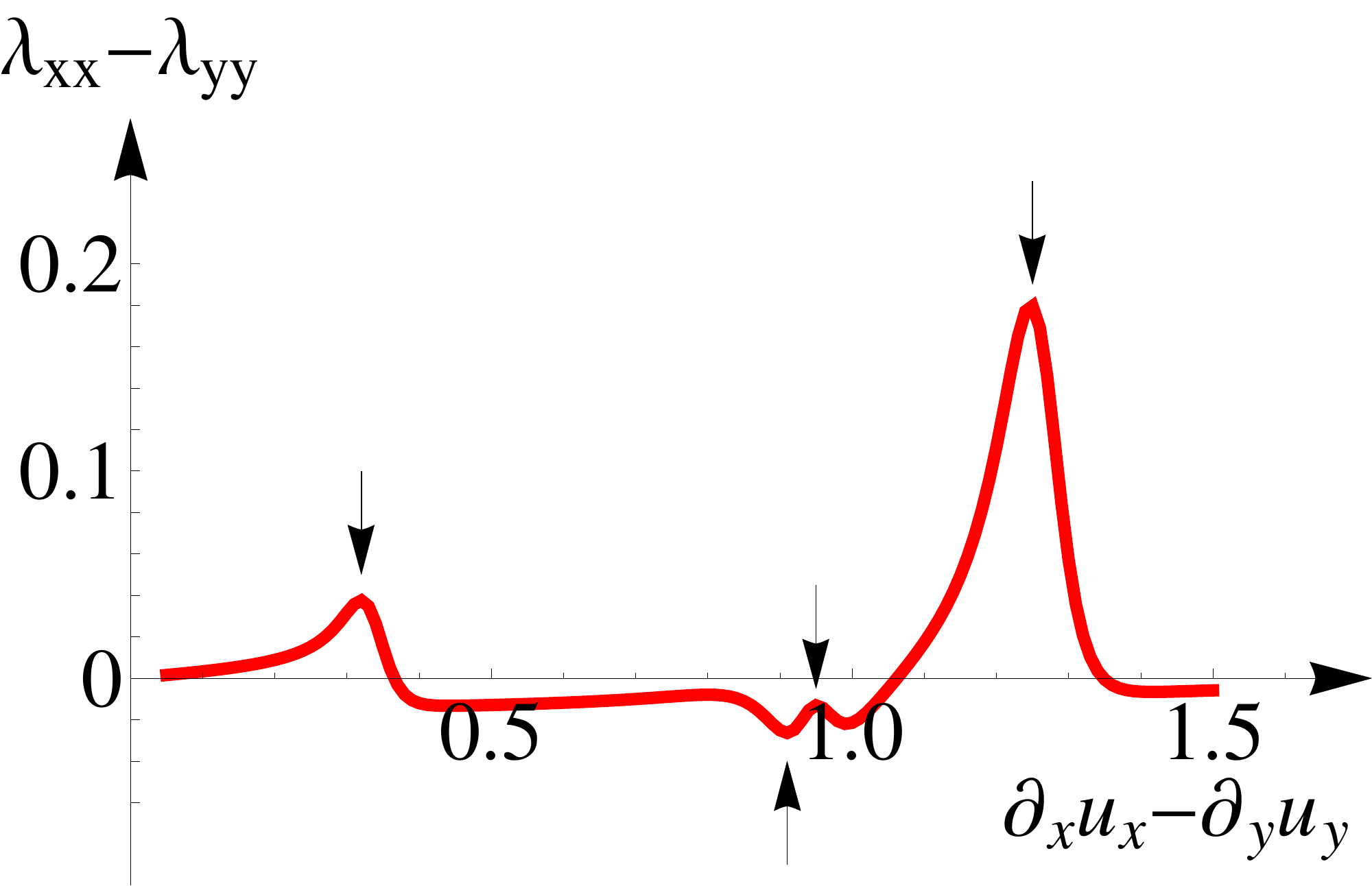}
\caption{Anisotropic penetration depth, $\lambda_{xx}-\lambda_{yy}$, as a
function of the mixing parameter $\gamma\propto\partial_{x}u_{x}-\partial_{y}u_{y}$
in Eq. (\ref{Eqn::SDGap}), for $T/\Delta_{0}=0.01$. Arrows denote
the peaks and troughs reflecting the annihilation and creation of
nodes in the Fermi pockets. \label{Fig:APD}}
\end{figure}

This behavior is more clearly manifested in anisotropic quantities,
such as the anisotropic penetration depth, $\lambda_{xx}-\lambda_{yy}$,
which can be measured by tunnel diode resonators/oscillators and microwave
cavities \cite{Matsuda_penetration_depth}. Formally, it is given
in terms of $\Delta\lambda_{\mu\mu}\propto-\int\mathrm{d}\omega\ \tilde{N}_{\mu\mu}(\omega)\frac{\partial f(\omega)}{\partial\omega}$,
where $f(\omega)$ is the Fermi-Dirac distribution function,$ $$\tilde{N}_{\mu\mu}(\omega)=\sum_{j}\int d^{2}k\ \delta(\omega-E_{j}(\bk))v_{j,\mu}^{F}v_{j,\mu}^{F}$
is the sum of the Fermi-velocity weighted density of states of all
$j$ pockets, and $E_{j}(\bk)$ is the corresponding quasi-particle
dispersion. Analysis of this quantity reveals that, at low temperatures,
the quadratic nodes only affect the component of the penetration depth
parallel to the direction in which they appear or disappear, i.e.
$\Delta\lambda_{\parallel}\propto\sqrt{T}$ whereas $\Delta\lambda_{\perp}\propto T$.
As a result, $\lambda_{\parallel}\gg\lambda_{\perp}$ when a quadratic
node appear or disappear from the Fermi surface, resulting in a peak
or a trough in $\lambda_{xx}-\lambda_{yy}$. To illustrate this behavior,
we plot in Fig. \ref{Fig:APD} $\lambda_{xx}-\lambda_{yy}$ as function
of uniaxial strain, using Eq. (\ref{Eqn::SDGap}) with $\gamma\propto\partial_{x}u_{x}-\partial_{y}u_{y}$,
and $\Delta_{s}$ and $\Delta_{d}$ obtained from the solution of
our effective model in Eq. (\ref{EQ_effective_model}) extended for
four bands (see supplementary material). Two large peaks, corresponding
to the annihilation of nodes in the electron pockets $e_{x}$ and
$e_{y}$, are observed at small and large strain values, respectively.
The peak resulting from the creation of nodes in the $h_{1}$ hole
pocket nearly cancels the trough associated with the creation of nodes
in $h_{2}$, as reflected by the weaker features in the intermediate
strain region. Thus, $\lambda_{xx}-\lambda_{yy}$ is a viable quantity
to probe the motion of nodes induced by uniaxial strain.

In summary, we showed that the multi-orbital character of the superconducting
state of the iron pnictides, allied to the presence of a large nematic
susceptibility, opens up the interesting possibility of enhancing
$T_{c}$ and manipulating gap nodes by uniaxial strain. Our results
rely on the proximity between the $s^{+-}$ and $d$-wave instabilities,
as predicted by several theoretical models \cite{Kuroki09,KemperNJP,Graser10,Ikeda10,Maiti11,Thomale11}
and recently reported by Raman experiments in certain compounds \cite{raman_mode1,raman_mode2}.
Our focus here was in systems that display accidental nodes already
in the tetragonal phase, which is the case for $\mathrm{BaFe_{2}(As_{1-x}P_{x})_{2}}$
and possibly for $\mathrm{FeSe}$, $\mathrm{Ba(Fe_{1-x}Ru_{x})_{2}As_{2}}$
and $\mathrm{KFe_{2}As_{2}}$. Similar arguments imply that nodeless
superconductors can be driven nodal by the application of uniaxial
strain -- in this case, however, nodes appear and disappear only in
one of the electron pockets. Interestingly, quantum criticality has
only been unambiguously detected in the nodal superconductor $\mathrm{BaFe_{2}(As_{1-x}P_{x})_{2}}$
\cite{Kasahara10,Matsuda_penetration_depth}, which begs the question
of whether nodal quasi-particles play a fundamental role in promoting
this behavior. The possible manipulation of nodes by external strain
may help shedding light on this important issue.

We thank C. Chen, A. Chubukov, H. Fu, S. Graser, P. Hirschfeld, T.
Maier, S. Maiti, A. Millis, R. Prozorov, M. Tanatar, X. Wang, and
V. Vakaryuk for helpful discussions.

\widetext
\vspace{0.5cm}
\begin{center}
\textbf{\large Supplementary for ``Manipulation of gap nodes by uniaxial strain
in iron-based superconductors''}
\end{center}
\setcounter{equation}{0}
\setcounter{figure}{0}
\setcounter{table}{0}
\makeatletter
\renewcommand{\theequation}{S\arabic{equation}}
\renewcommand{\thefigure}{S\arabic{figure}}


\section{Effective three band model}

Our effective three band model contains the inner hole pocket $h_{1}$
and the two electron pockets $e_{x}$ and $e_{y}$. The outer hole
pocket is ignored since the RPA-derived pairing interaction connecting
$h_{2}$ to $e_{x/y}$ is weaker \cite{KemperNJP}. As discussed in
the main text, the (normalized) wave-function of the three Fermi surface
pockets is expanded in harmonic functions of the angles around the
pockets (measured with respect to the $x$ axis):
\begin{equation}
\begin{aligned}|h_{1}\rangle & =\dfrac{\cos\theta(1+\beta_{h}\sin^{2}\theta)|d_{xz}\rangle+\sin\theta(1-\beta_{h}\cos^{2}\theta)|d_{yz}\rangle}{\left(1+\beta_{h}^{2}\sin^{2}\theta\cos^{2}\theta\right)^{1/2}}\\
|e_{x}\rangle & =\dfrac{\alpha|d_{xy}\rangle+(1+\beta_{e})\sin\phi_{x}|d_{yz}\rangle}{\left(\alpha^{2}+(1+\beta_{e})^{2}\sin^{2}\phi_{x}\right)^{1/2}}\\
|e_{y}\rangle & =\dfrac{\alpha|d_{xy}\rangle+(1-\beta_{e})\cos\phi_{y}|d_{yz}\rangle}{\left(\alpha^{2}+(1-\beta_{e})^{2}\cos^{2}\phi_{y}\right)^{1/2}}\ ,
\end{aligned}
\label{Eqn::OrbitalFit}
\end{equation}

\begin{figure}[htbp]
\centering \subfigure[\label{Fig:OrbitalWeight:Ex}]{\includegraphics[width=3in]{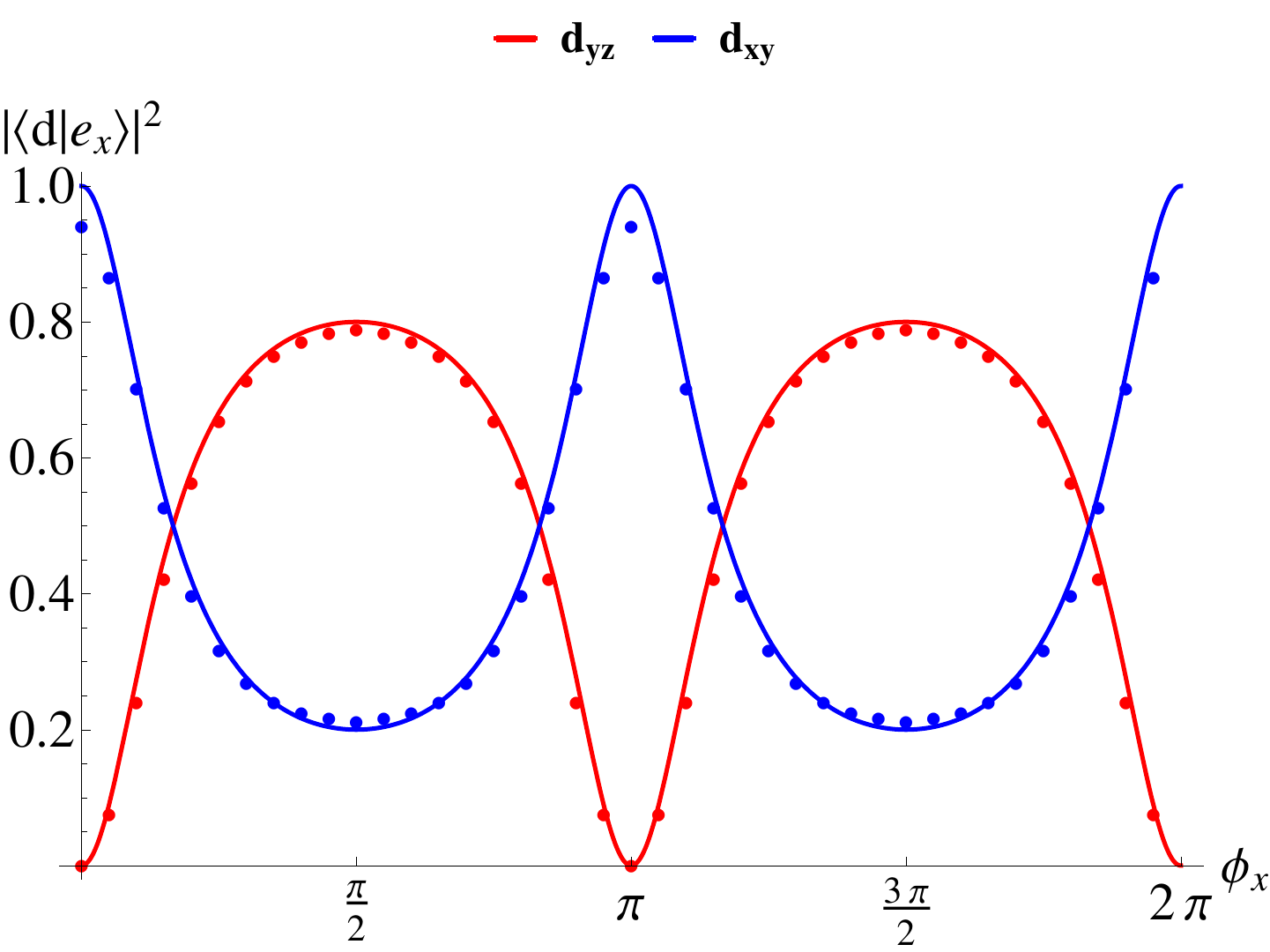}}
\subfigure[\label{Fig:OrbitalWeight:Hole}]{\includegraphics[width=3in]{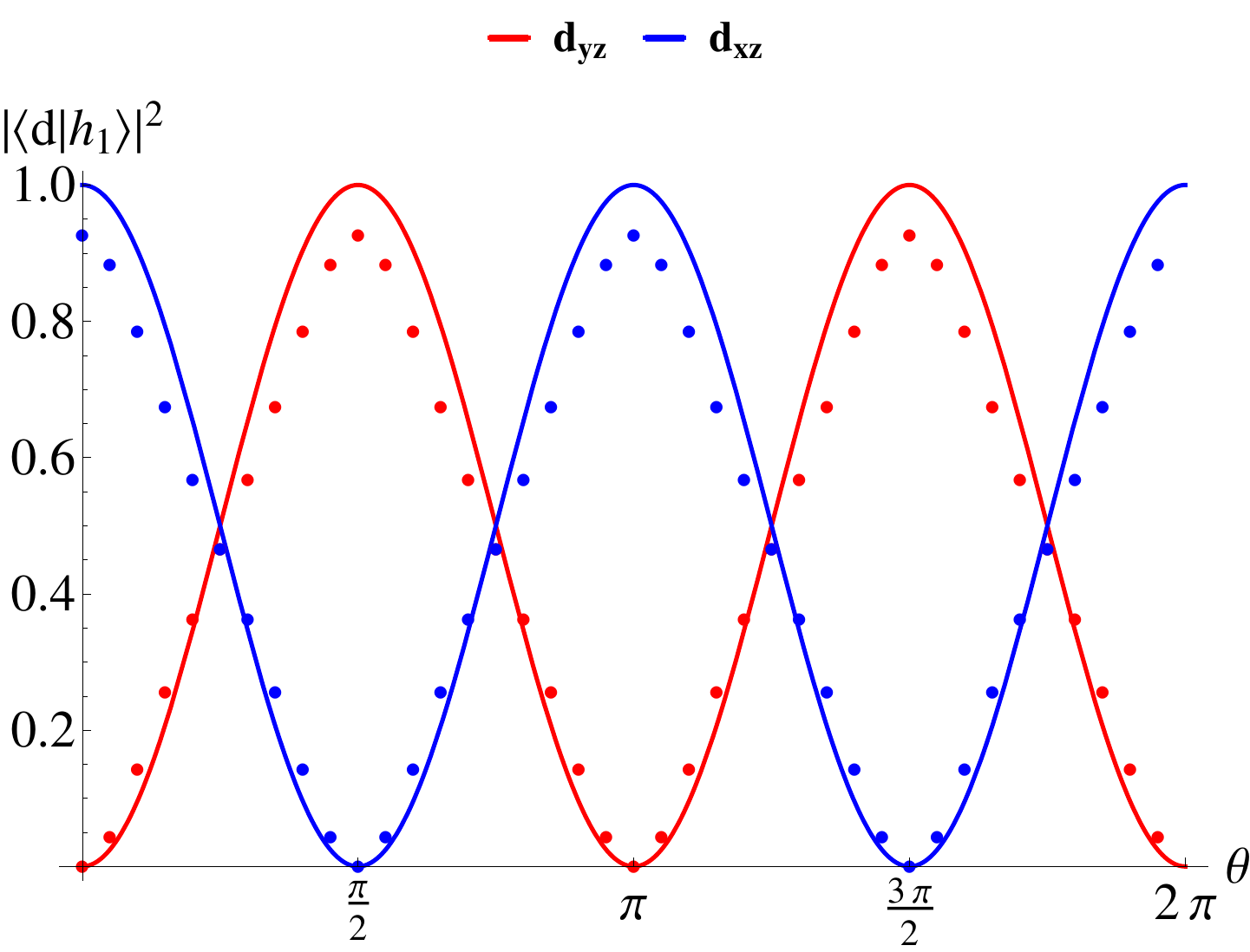}}
\caption{Orbital weight derived from the five orbital tight binding model (dots)
and from the effective model, Eq. \eqref{Eqn::OrbitalFit}, for $\alpha=0.52$.
(a) $d_{xy}$ (blue) and $d_{yz}$ (red) orbital weights on the $e_{x}$
pocket. (b) $d_{yz}$ (red) and $d_{xz}$ (blue) orbital weights on
the inner hole pocket $h_{1}$. }

\label{Fig:OrbitalWeight}
\end{figure}

Here, $\alpha$ controls the angular dependence of the orbital weights
on the electron pockets in the tetragonal phase, and $\beta_{h,e}$
describe the change in the orbital weights promoted by orbital order.
To focus on the effects caused by the changes in orbital weight, and
since the modification in the shapes of the Fermi pockets are small,
we consider for simplicity three equal circular pockets. To obtain
$\alpha$, we fit the matrix elements $\left\langle \left.d_{j}\right|e_{x}\right\rangle $
obtained from the equation above to those obtained from the tight-binding
model. We find that $\alpha\approx0.52$ yields a satisfactory description,
as shown in Fig. \ref{Fig:OrbitalWeight}.

To obtain the pairing interaction in the band basis, we consider only
the intra-orbital contributions:
\begin{align}
\Gamma_{ij}\left(\mathbf{k},\mathbf{k}'\right) & =\sum_{i}\langle i|m\rangle_{\mathbf{k}}\langle i|m\rangle_{-\mathbf{k}}\ \Gamma_{mm}^{mm}\ \langle m|j\rangle_{\mathbf{k}'}\langle m|j\rangle_{-\mathbf{k}'}\ ,
\end{align}
 where $i$ and $j$ are band indices and $m$ is the orbital index.
$\Gamma_{mm}^{mm}$ is the momentum independent pairing interaction
in the orbital basis. Based on the RPA-derived pairing interaction,
we consider only the intra-orbital pairing interactions connecting
the same orbital on different pockets, $\Gamma_{mm}^{mm}=V$ for $m=d_{xz/yz}$
and $\Gamma_{mm}^{mm}=W$ for $m=d_{xy}$. Therefore:
\begin{align}
\Gamma_{xy} & =\dfrac{\alpha^{4}V}{\left(\alpha^{2}+(1+\beta_{e})^{2}\cos^{2}\phi_{x}\right)\left(\alpha^{2}+(1-\beta_{e})^{2}\cos^{2}\phi_{y}\right)}\\
\Gamma_{hx} & =\dfrac{\sin^{2}\theta(1-\beta_{h}\cos^{2}\theta)^{2}}{1+\beta_{h}^{2}\sin^{2}\theta\cos^{2}\theta}\dfrac{(1+\beta_{e})^{2}\sin^{2}\phi_{x}}{\alpha^{2}+(1+\beta_{e})^{2}\sin^{2}\phi_{x}}V\\
\Gamma_{hy} & =\dfrac{\cos^{2}\theta(1+\beta_{h}\sin^{2}\theta)^{2}}{1+\beta_{h}^{2}\sin^{2}\theta\cos^{2}\theta}\dfrac{(1-\beta_{e})^{2}\cos^{2}\phi_{y}}{\alpha^{2}+(1-\beta_{e})^{2}\cos^{2}\phi_{y}}W
\end{align}

It is then straightforward to write down the BCS-like gap equations
at an arbitrary temperature $T$. Denoting $v=N_{0}V$ and $w=N_{0}W$,
where $N_{0}$ is the density of states at the Fermi level, we obtain:
\begin{align}
-\Delta_{h}(\theta) & =v\,\dfrac{\sin^{2}\theta(1-\beta_{h}\cos^{2}\theta)^{2}}{1+\beta_{h}^{2}\sin^{2}\theta\cos^{2}\theta}\int\frac{\dif\phi_{x}}{2\pi}\dfrac{\Delta_{x}(\phi_{x})\,(1+\beta_{e})^{2}\sin^{2}\phi_{x}}{\alpha^{2}+(1+\beta_{e})^{2}\sin^{2}\phi_{x}}\int_{0}^{\Lambda}d\varepsilon\frac{\tanh\left(\frac{\sqrt{\varepsilon^{2}+\Delta_{x}^{2}(\phi_{x})}}{2T}\right)}{\sqrt{\varepsilon^{2}+\Delta_{x}^{2}(\phi_{x})}}\nonumber \\
 & +v\,\dfrac{\cos^{2}\theta(1+\beta_{h}\sin^{2}\theta)^{2}}{1+\beta_{h}^{2}\sin^{2}\theta\cos^{2}\theta}\int\frac{\dif\phi_{y}}{2\pi}\dfrac{\Delta_{y}(\phi_{y})\,(1-\beta_{e})^{2}\sin^{2}\phi_{y}}{\alpha^{2}+(1-\beta_{e})^{2}\sin^{2}\phi_{y}}\int_{0}^{\Lambda}d\varepsilon\frac{\tanh\left(\frac{\sqrt{\varepsilon^{2}+\Delta_{y}^{2}(\phi_{y})}}{2T}\right)}{\sqrt{\varepsilon^{2}+\Delta_{y}^{2}(\phi_{y})}}\label{Eqn:SCHoleOO}\\
-\Delta_{x}(\phi_{x}) & =\dfrac{v\,(1+\beta_{e})^{2}\sin^{2}\phi_{x}}{\alpha^{2}+(1+\beta_{e})^{2}\sin^{2}\phi_{x}}\int\frac{\dif\theta}{2\pi}\dfrac{\Delta_{h}(\theta)\,\sin^{2}\theta(1-\beta_{h}\cos^{2}\theta)^{2}}{1+\beta_{h}\sin^{2}\theta\cos^{2}\theta}\int_{0}^{\Lambda}d\varepsilon\frac{\tanh\left(\frac{\sqrt{\varepsilon^{2}+\Delta_{h}^{2}(\theta)}}{2T}\right)}{\sqrt{\varepsilon^{2}+\Delta_{h}^{2}(\theta)}}\nonumber \\
 & +\dfrac{w\,\alpha^{4}}{\alpha^{2}+(1+\beta_{e})^{2}\sin^{2}\phi_{x}}\int\frac{\dif\phi_{y}}{2\pi}\dfrac{\Delta_{y}(\phi_{y})}{\alpha^{2}+(1-\beta_{e})^{2}\sin^{2}\phi_{y}}\int_{0}^{\Lambda}d\varepsilon\frac{\tanh\left(\frac{\sqrt{\varepsilon^{2}+\Delta_{y}^{2}(\phi_{y})}}{2T}\right)}{\sqrt{\varepsilon^{2}+\Delta_{y}^{2}(\phi_{y})}}\label{Eqn:SCExOO}\\
-\Delta_{y}(\phi_{y}) & =\dfrac{v\,(1-\beta_{e})^{2}\sin^{2}\phi_{y}}{\alpha^{2}+(1-\beta_{e})^{2}\sin^{2}\phi_{y}}\int\frac{\dif\theta}{2\pi}\dfrac{\Delta_{h}(\theta)\,\cos^{2}\theta(1+\beta_{h}\sin^{2}\theta)^{2}}{1+\beta_{h}^{2}\sin^{2}\theta\cos^{2}\theta}\int_{0}^{\Lambda}d\varepsilon\frac{\tanh\left(\frac{\sqrt{\varepsilon^{2}+\Delta_{h}^{2}(\theta)}}{2T}\right)}{\sqrt{\varepsilon^{2}+\Delta_{h}^{2}(\theta)}}\nonumber \\
 & +\dfrac{w\,\alpha^{4}}{\alpha^{2}+(1-\beta_{e})^{2}\sin^{2}\phi_{y}}\int\frac{\dif\phi_{y}}{2\pi}\dfrac{\Delta_{x}(\phi_{x})}{\alpha^{2}+(1+\beta_{e})^{2}\sin^{2}\phi_{x}}\int_{0}^{\Lambda}d\varepsilon\frac{\tanh\left(\frac{\sqrt{\varepsilon^{2}+\Delta_{x}^{2}(\phi_{x})}}{2T}\right)}{\sqrt{\varepsilon^{2}+\Delta_{x}^{2}(\phi_{x})}}\label{Eqn:SCEyOO}
\end{align}
 where $\Lambda$ is the upper cutoff. To solve them at $T=T_{c}$
and at $T=0$, it is convenient to also expand the gaps in harmonic
functions:

\begin{align}
\Delta_{h}(\theta) & =\dfrac{c_{0}+c_{2}\cos2\theta+c_{4}\cos4\theta+c_{6}\cos6\theta}{1+\beta_{h}^{2}\sin^{2}\theta\cos^{2}\theta}\nonumber \\
\Delta_{x}(\phi_{x}) & =\frac{a_{x}+b_{x}\sin^{2}\phi_{x}}{\alpha^{2}+(1+\beta_{e})^{2}\sin^{2}\phi_{x}}\qquad\Delta_{y}(\phi_{y})=\frac{a_{y}+b_{y}\cos^{2}\phi_{y}}{\alpha^{2}+(1-\beta_{e})^{2}\cos^{2}\phi_{y}}\ .\label{expansion}
\end{align}

\begin{figure}[htbp]
\centering \subfigure[\label{Fig:NoOOSC:Eigen}]{\includegraphics[width=3in]{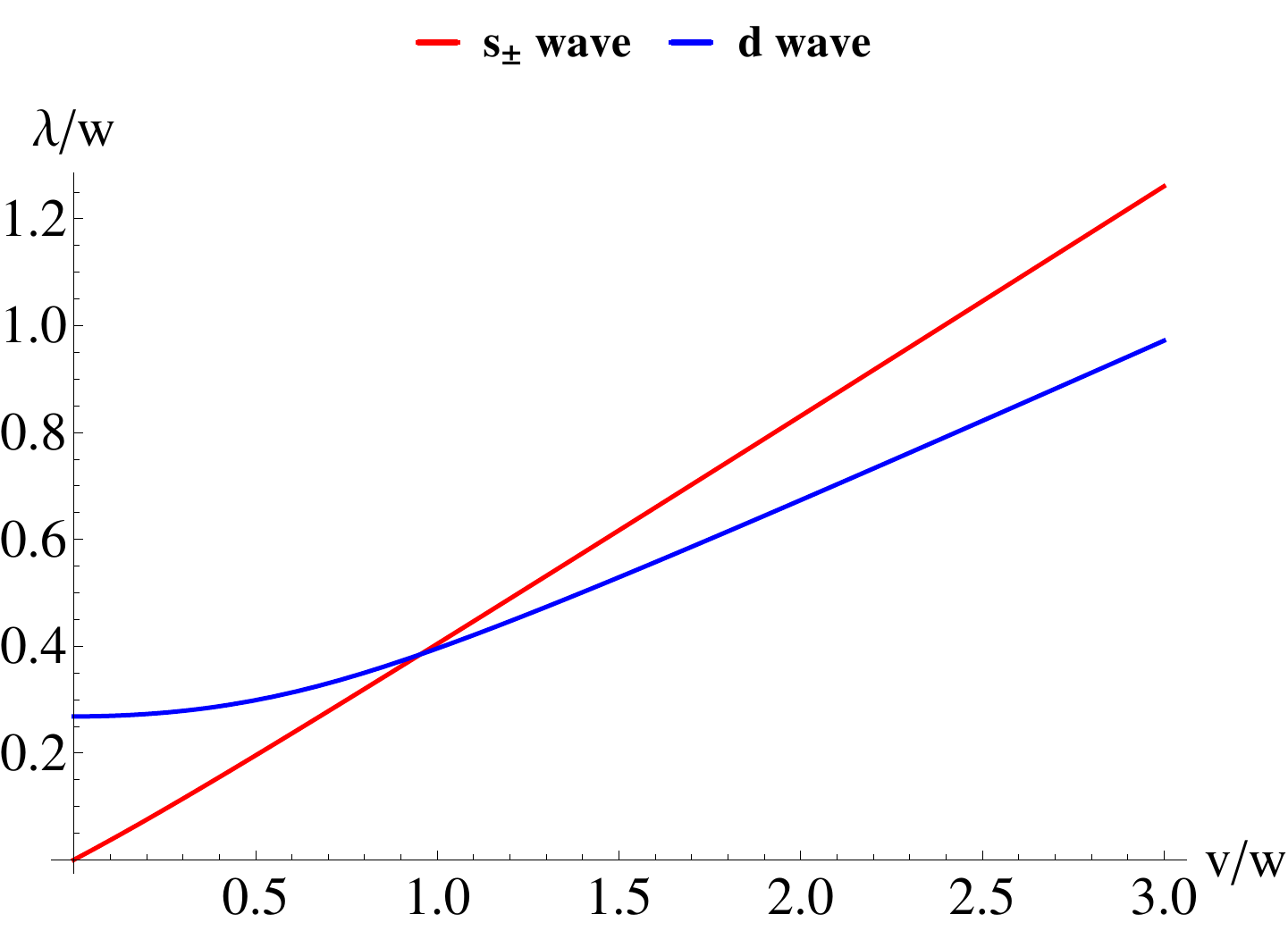}}
\subfigure[\label{Fig:NoOOSC:Node}]{\includegraphics[width=3in]{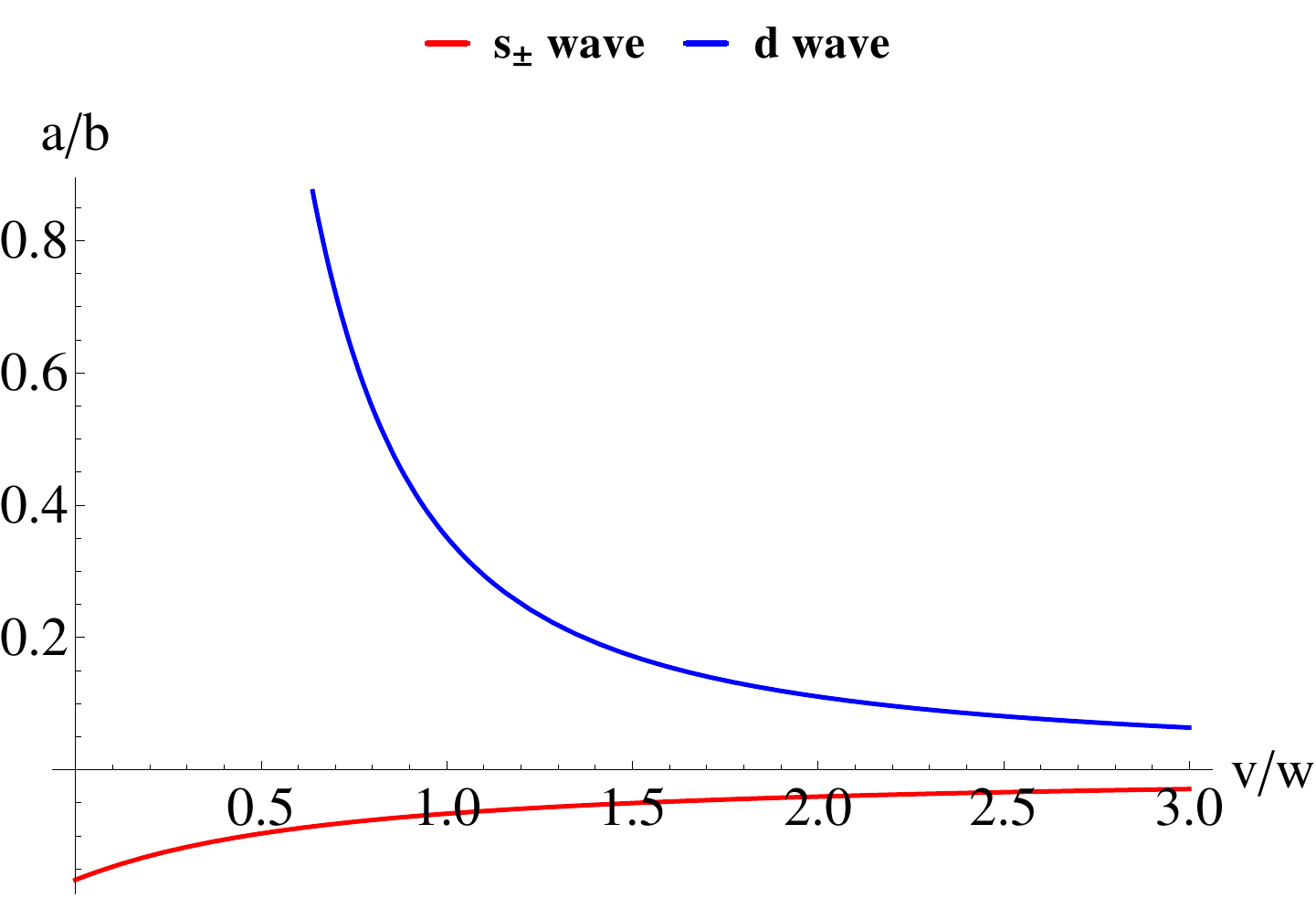}}
\caption{(a) Eigenvalue $\lambda$ of different superconducting states as a
function of the ratio $v/w$ between intra-orbital pairing involving
$d_{xz/yz}$ orbitals and $d_{xy}$ orbitals, respectively. (b) Ratio
of the coefficients $a_{x}$ and $b_{x}$ of the gap in the electron
pocket $e_{x}$, given in Eq. (\ref{expansion}), as function of $v/w$. }

\label{Fig:NoOOSC}
\end{figure}

At $T=T_{c}$, the gap equations are then reduced to an $8\times8$
linear system of equations in the coefficients $c_{i}$, $a_{i}$,
and $b_{i}$, and the leading instability is the one with the largest
eigenvalue $\lambda=\ln\left(\frac{1.13\Lambda}{T_{c}}\right)$. In
the absence of orbital order, $\beta_{h}=\beta_{e}=0$, the leading
solutions are the $s^{+-}$ state (for $v\gtrsim w$) and the $d$-wave
state (for $v\lesssim w$), as shown in Fig.~\ref{Fig:NoOOSC}. In
the $s^{+-}$ state, $a_{y}=a_{x}$, $b_{y}=b_{x}$, $c_{2}=c_{4}=c_{6}=0$,
and $\mathrm{sign}\: c_{0}=-\mathrm{sign\left[a+b\left(-\alpha^{2}+\alpha\sqrt{1+\alpha^{2}}\right)\right]}$.
We find $-1<a_{x}/b_{x}<0$, implying that accidental nodes appear
in the electron pockets. Because $c_{2}=c_{4}=c_{6}=0$, accidental
nodes do not appear in the hole pocket. In the $d$-wave state, $a_{y}=-a_{x}$,
$b_{y}=-b_{x}$, and $c_{0}=c_{4}=c_{6}=0$. Our solution gives $a_{x}/b_{x}>0$,
which precludes nodes from appearing in the electron pockets. As expected
by symmetry, nodes appear in the hole pocket at $\theta=\pm\pi/4$
and $\theta=\pm3\pi/4$.

\begin{figure}[htbp]
\centering \subfigure[\label{Fig:OOSC:Eigen}]{\includegraphics[width=2.5in]{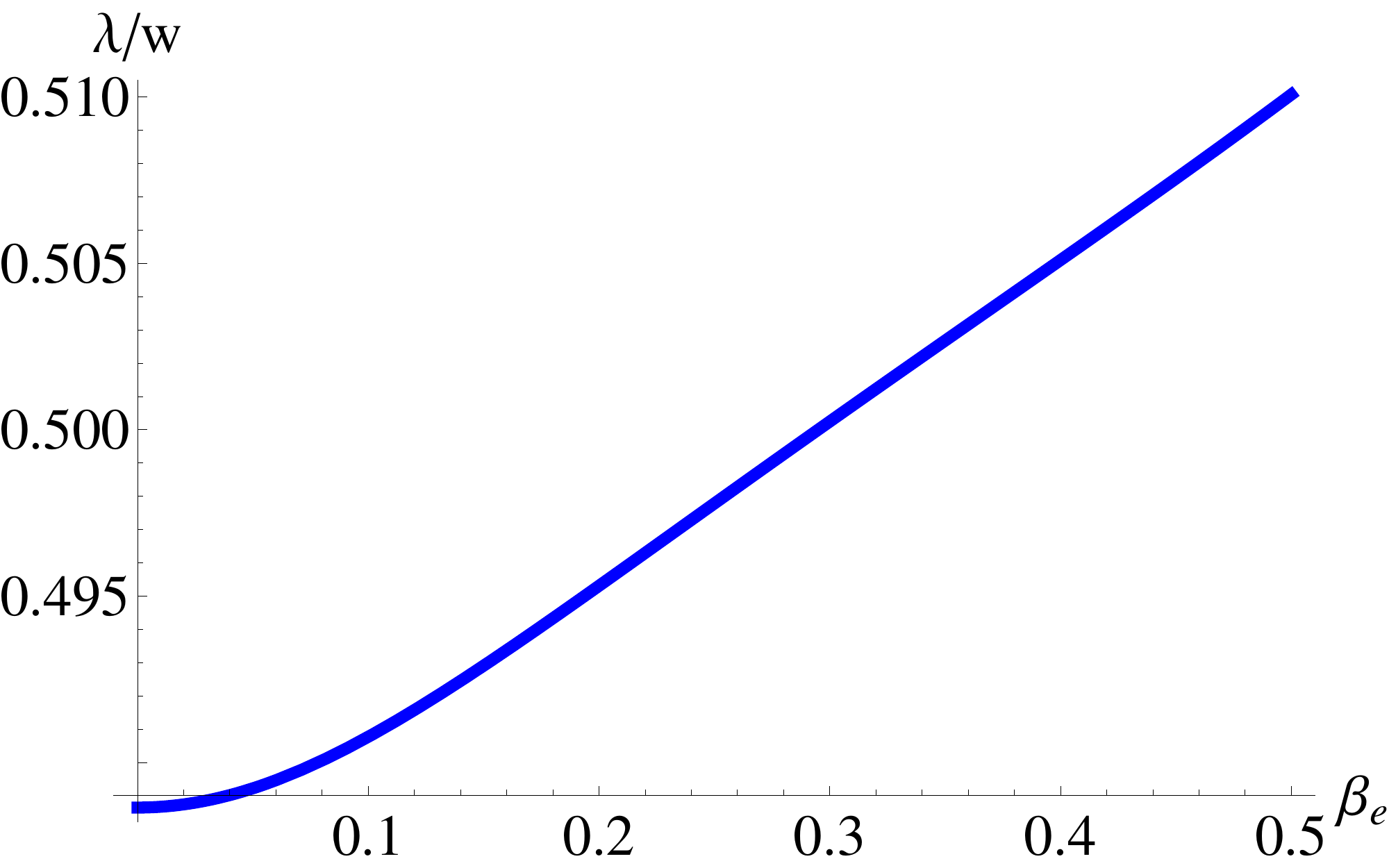}}
\subfigure[\label{Fig:OOSC:Node}]{\includegraphics[width=2.5in]{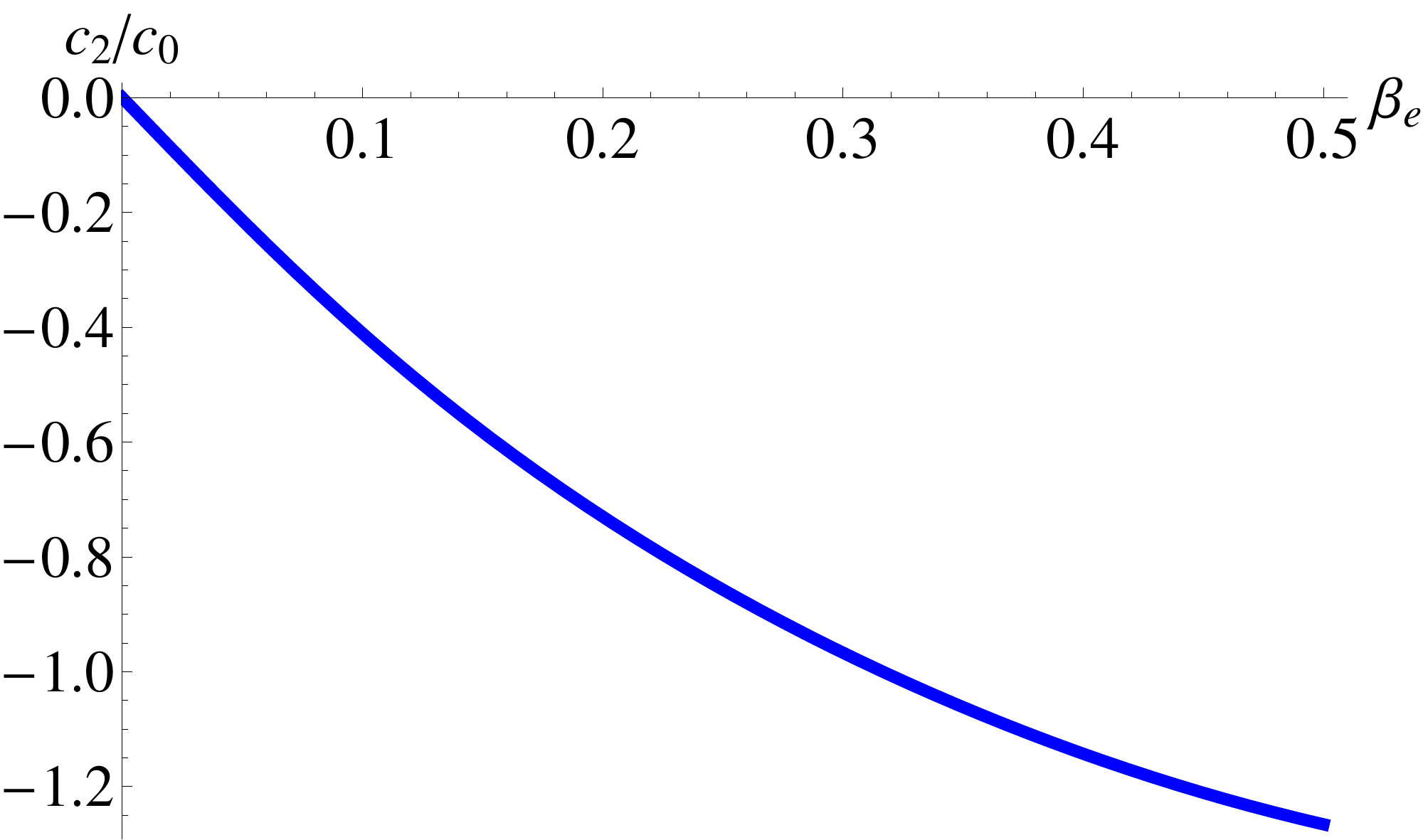}}
\subfigure[\label{Fig:OOSC:ExNode}]{\includegraphics[width=2.5in]{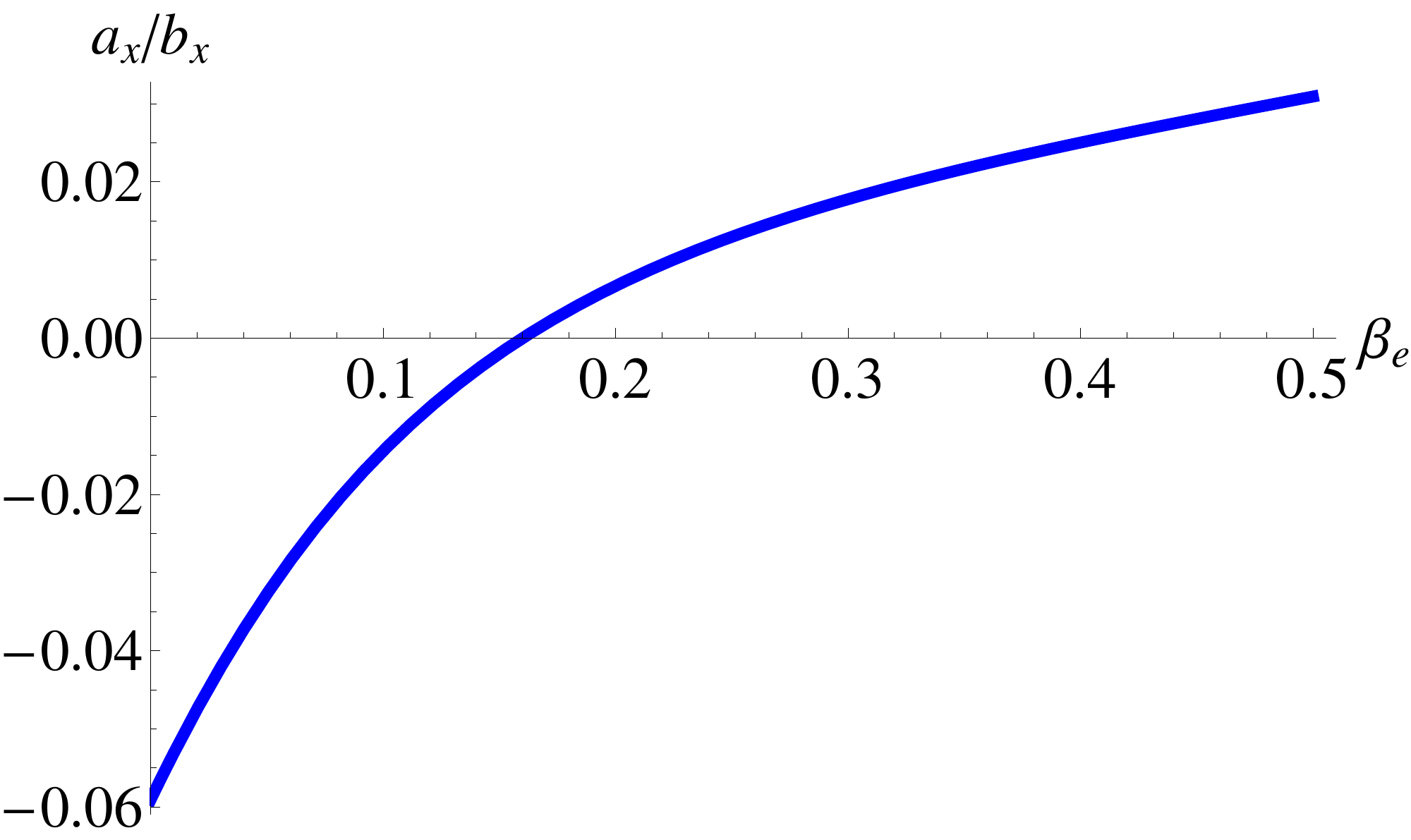}}
\subfigure[\label{Fig:OOSC:EyNode}]{\includegraphics[width=2.5in]{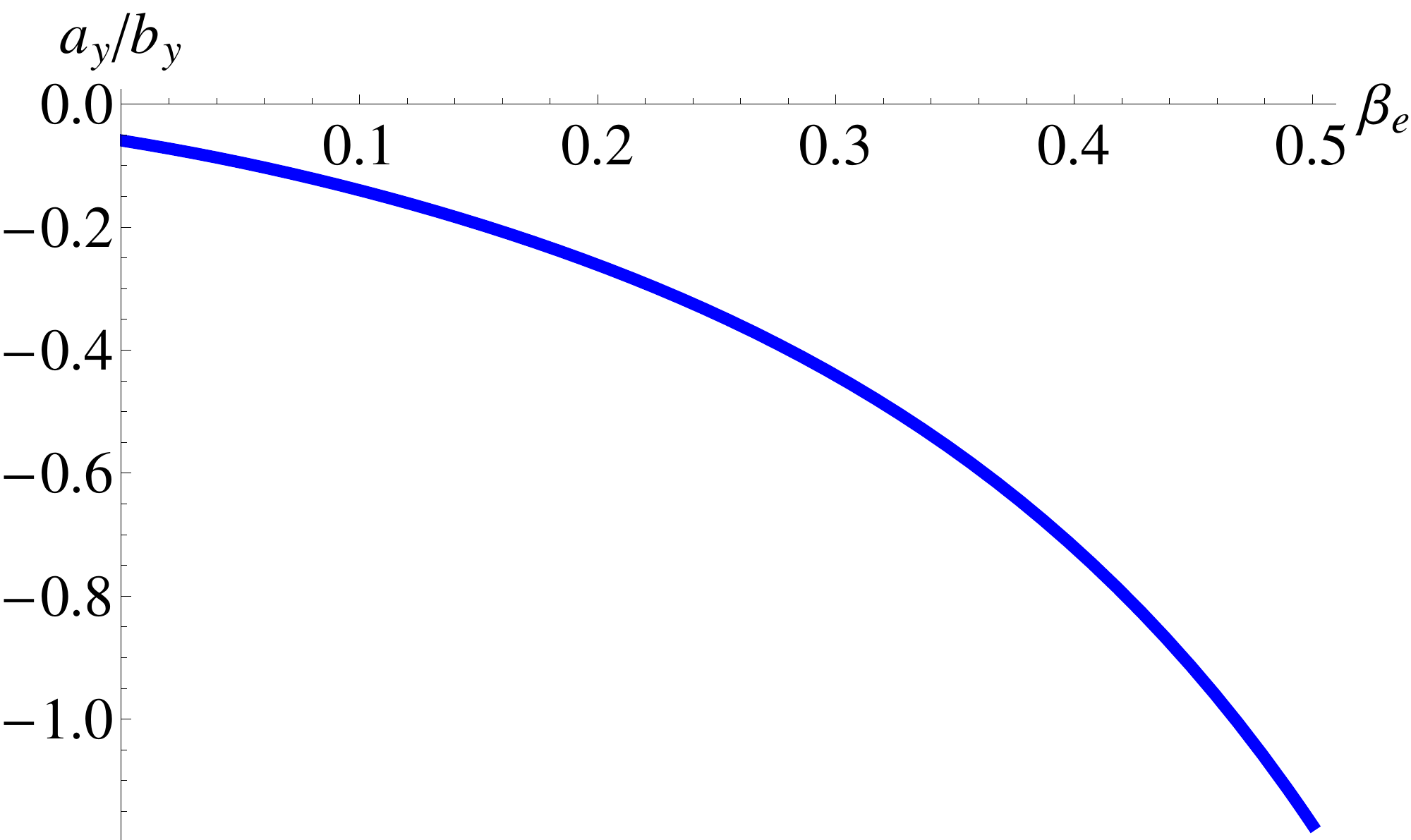}}
\caption{In the plots above, we set $\beta_{h}=\beta_{e}/2$ and $v/w=1.2$,
placing the system near the degeneracy between $s^{+-}$ and $d$-wave.
(a) Leading superconducting eigenvalue as a function of $\beta_{e}$.
(b) Ratio of the leading coefficients $c_{0}$ and $c_{2}$ of the
gap in the hole pocket $h_{1}$, given in Eq. (\ref{expansion}),
as function of $\beta_{e}$. (c) Ratio of the coefficients $a_{x}$
and $b_{x}$ of the gap in the electron pocket $e_{x}$, given in
Eq. (\ref{expansion}), as function of $\beta_{e}$. (d) Ratio of
the coefficients $a_{y}$ and $b_{y}$ of the gap in the electron
pocket $e_{y}$, given in Eq. (\ref{expansion}), as function of $\beta_{e}$.}

\label{Fig:OOSC}
\end{figure}

In the presence of orbital order, both $\beta_{e}$ and $\beta_{h}$
become non-zero. To reflect the fact that the RPA-derived pairing
interaction between the hole pocket and the $e_{x}$ pocket is strongly
enhanced by orbital order, we must set $\beta_{e}>\beta_{h}$. For
concreteness, here we consider $\beta_{e}=2\beta_{h}$. Solving the
linearized gap equations as function of $\beta_{e}$ we find an increase
in the leading eigenvalue $\lambda$ when $ $$w$ and $v$ are comparable,
$v=1.2w$, (i.e. nearly-degenerate $s^{+-}$ and $d$-wave states),
as shown in Fig.~\ref{Fig:OOSC}. Since $T_{c}\propto\mathrm{e}^{-1/\lambda}$,
$T_{c}$ is enhanced by orbital order, in agreement with the RPA calculations.

As for the gap structure, we find that nodes are introduced in the
hole pocket along the $x$ direction for $\beta_{e}\gtrsim0.36$,
when $c_{2}/c_{0}<-1$. In both electron pockets, nodes are expelled
along the $x$ axis -- in particular, they are expelled from $e_{x}$
for $\beta_{e}\gtrsim0.16$ (when $a_{x}/b_{x}>0$) and from the $e_{y}$
pocket for $\beta_{e}\gtrsim0.45$ (when $a_{y}/b_{y}<-1$). The fact
that the nodes are expelled from the $e_{x}$ pocket before they leave
the $e_{y}$ pocket also agrees with the RPA calculations.

Having established a model that captures the main features of the
RPA calculation at $T_{c}$, we calculated the gap structure also
for $T=0$ to check whether the nodal structure is robust with changes
in temperature. We find that, independent of the value of the cutoff
$\Lambda$, the gap structure at $T=0$ is very similar to the one
at $T_{c}$, as shown in Fig.~3 of the manuscript.

\section{Anisotropic penetration depth}

\subsection{Scaling behavior of quadratic nodes}

The penetration depth measured along the $\mu$ direction for a field
applied in the $\mu$ direction is given by:

\begin{equation}
\Delta\lambda_{\mu\mu}\equiv\lambda_{\mu\mu}-\lambda_{0}=\alpha\int\mathrm{d}\omega\ \tilde{N}_{\mu\mu}(\omega)\left(-\frac{\partial f(\omega)}{\partial\omega}\right)\label{S_Delta_lambda}
\end{equation}
 where $\lambda_{0}$ is the $T=0$ value of the penetration depth,
$\alpha$ is an overall pre-factor, $f\left(\omega\right)$ is the
Fermi-Dirac distribution function, and $\tilde{N}\left(\omega\right)$
is the Fermi-velocity weighted density of states:

\begin{equation}
\tilde{N}_{\mu\mu}(\omega)=\sum_{j}\int d^{2}k\ \delta\left(\omega-E_{j}(\mathbf{k})\right)\left(v_{j,\mu}^{F}\right)^{2}\label{S_N}
\end{equation}

Here, $E_{j}\left(\mathbf{k}\right)=\sqrt{\Delta_{j}^{2}\left(\mathbf{k}\right)+\varepsilon^{2}\left(\mathbf{k}\right)}$
is the quasi-particle dispersion of pocket $j$, with $\varepsilon\left(\mathbf{k}\right)=\frac{k^{2}}{2m}-\varepsilon_{F}$
assumed to be a parabolic dispersion. Note that, to focus on the effect
caused by the changes in the gap function due to orbital order, we
assume the four pockets to be the same -- this simplification does
not affect the main results below.

Let us consider the effect of a quadratic node on a single pocket.
The quadratic node appears when the node is on the verge of being
introduced in or expelled from the pocket. For concreteness, we consider
the gap function on the electron pocket $j$:

\begin{equation}
\Delta_{j}\left(\phi_{j}\right)=\Delta_{0,j}\left(1-r\cos2\phi_{j}\right)\label{accidental_nodes}
\end{equation}

For the hole pockets, one only needs to change $\cos2\phi_{j}$ to
$\cos4\phi_{j}$, and the results are similar. To avoid cumbersome
expressions, we drop the subscript $j$ hereafter. Note that for $\left|r\right|>1$,
accidental nodes are present, whereas for $\left|r\right|<1$, nodes
are absent. Thus $r=\pm1$ corresponds to the point where nodes disappear
from (or appear in) the pocket. For $r=1$, the nodes are expelled
along the $x$ direction ($\phi_{0}=0,\pi$), whereas for $r=-1$,
they are expelled along the $y$ direction ($\phi_{0}=\pm\pi/2$).
Expansion around either of these points gives $\Delta\left(\phi\right)\propto\left(\phi-\phi_{0}\right)^{2}$,
hence the name quadratic nodes.

To illustrate the low-temperature scaling behavior of the anisotropic
penetration depth, we consider the case $r=1$. Substituting in Eq.
(\ref{S_N}) we obtain:

\begin{align}
\tilde{N}_{xx}(\omega)= & N_{0}v_{F}^{2}\omega\int\frac{d\phi}{2\pi}\ \frac{\cos^{2}\phi}{\sqrt{\omega^{2}-\Delta_{0}^{2}\left(1-\cos2\phi\right)^{2}}}\theta\left(\omega-\Delta_{0}\left(1-\cos2\phi\right)\right)\nonumber \\
\tilde{N}_{yy}(\omega)= & N_{0}v_{F}^{2}\omega\int\frac{d\phi}{2\pi}\ \frac{\sin^{2}\phi}{\sqrt{\omega^{2}-\Delta_{0}^{2}\left(1-\cos2\phi\right)^{2}}}\theta\left(\omega-\Delta_{0}\left(1-\cos2\phi\right)\right)\label{aux_N1}
\end{align}
 where $\theta\left(x\right)$ is the step function. For $\omega\ll\Delta_{0}$,
we can expand around the quadratic node $\phi_{0}=0$ to obtain:

\begin{align}
\tilde{N}_{xx}(\omega)= & N_{0}v_{F}^{2}\sqrt{\frac{\omega}{2\Delta_{0}}}\int_{-1}^{1}\frac{dz}{2\pi}\ \frac{1}{\sqrt{1-z^{4}}}\sim\sqrt{\omega}\nonumber \\
\tilde{N}_{yy}(\omega)= & N_{0}v_{F}^{2}\left(\frac{\omega}{2\Delta_{0}}\right)^{3/2}\int_{-1}^{1}\frac{dz}{2\pi}\ \frac{z^{2}}{\sqrt{1-z^{4}}}\sim\omega^{3/2}\label{aux_N2}
\end{align}

Substitution in Eq. (\ref{S_Delta_lambda}) then gives the low-temperature
behavior $\Delta\lambda_{xx}\sim\sqrt{T}$ and $\Delta\lambda_{yy}\sim T^{3/2}$.
Note that, if accidental linear nodes are present in other pockets,
they will give a linear-in-$T$ contribution $\Delta\lambda_{\mu\mu}\sim T$
for both directions. Thus, while the quadratic nodes dominate the
low-temperature behavior of the penetration depth along the $x$ direction,
accidental linear nodes dominate the behavior of the penetration depth
along the $y$ direction. Note also that for $r=-1$, when nodes are
expelled from the pocket along the $y$ direction, one obtains the
opposite behavior $\Delta\lambda_{xx}\sim T^{3/2}$ and $\Delta\lambda_{yy}\sim\sqrt{T}$.

\subsection{Four-band model}

To calculate how the anisotropic penetration depth changes as orbital
order increases, we start with the expression that relates the gap
structure in the presence of orbital order to a mixture of the $s^{+-}$
and $d$-wave gap structures of the tetragonal phase:

\begin{equation}
\Delta_{i}=\Delta_{i,s}+\gamma\Delta_{i,d}\label{gap_mixture}
\end{equation}
 where $i$ denotes one of the four Fermi pockets. As discussed in
the main text, increasing orbital order implies increasing the mixing
coupling $\gamma$. Since the nodal structure at $T_{c}$ is very
similar to the nodal structure at $T=0$ (Fig. 3 of the main paper),
we write the gaps as the product of an overall temperature-dependent
(but pocket-independent) amplitude $\Delta_{0}$ and a temperature-independent
structure factor $g_{i,s/d}$, i.e. $\Delta_{i,s/d}=\Delta_{0}\, g_{i,s/d}$.
Thus, we can obtain $g_{i,s/d}$ directly from the linearized gap
equations in the tetragonal phase.

To capture the effects of the nodes that emerge in the outer hole
pocket $h_{2}$, we expand our effective three-band model by including
the normalized wave-function:
\begin{equation}
|h_{2}\rangle=\sin\theta\left|d_{xz}\right\rangle -\cos\theta\left|d_{yz}\right\rangle \ .\label{h2}
\end{equation}
 such that $ $$\left\langle \left.h_{2}\right|h_{1}\right\rangle =0$.
Note that, in accord to the tight-binding model, the angular variation
of the orbital content of $h_{2}$ is $\pi/2$ out of phase with respect
to the angular variation of the orbital content of $h_{1}$. Since
we will solve the linearized gap equations in the tetragonal phase,
i.e. $\beta_{h}=\beta_{e}=0$, the pairing interactions in the band
basis acquire the simplified form:
\begin{align}
\Gamma_{xy} & =\frac{\alpha^{4}W}{(\alpha^{2}+\sin^{2}\phi_{x})(\alpha^{2}+\cos^{2}\phi_{y})}\\
\Gamma_{h_{1}x} & =\frac{V\sin^{2}\phi_{x}\sin^{2}\theta}{\alpha^{2}+\sin^{2}\phi_{x}} & \Gamma_{h_{1}y} & =\frac{V\cos^{2}\phi_{y}\cos^{2}\theta}{\alpha^{2}+\cos^{2}\phi_{y}}\\
\Gamma_{h_{2}x} & =t\frac{V\sin^{2}\phi_{x}\cos^{2}\theta}{\alpha^{2}+\sin^{2}\phi_{x}} & \Gamma_{h_{2}y} & =t\frac{V\cos^{2}\phi_{y}\sin^{2}\theta}{\alpha^{2}+\cos^{2}\phi_{y}}\ .
\end{align}

The RPA calculation reveals that the pairing interaction is greatly
enhanced by the nesting between $h_{1}$ and the electron pockets
\cite{KemperNJP}. To account for the worse nesting conditions between
$h_{2}$ and the electron pockets, we add a factor $t<1$ to the pairing
interactions $\Gamma_{h_{2}x/y}$. For concreteness, here we set $t=2/3$.

The linearized gap equation can be conveniently expressed as an algebraic
system of equations after writing the gap functions in terms of harmonic
functions of the angle around the pockets, Eq. (\ref{expansion}).
Note that, because we are in the tetragonal state, $c_{4}=c_{6}=0$
always. Solution of the gap equations reveals that the phase diagram
is very similar to the three-band model, with the $s^{+-}$ and $d$-wave
states becoming degenerate at $v/w\approx0.8$.

To calculate the anisotropic penetration depth, we place the system
near this degeneracy point -- in particular, we set $v=w$. At $T=T_{c}$,
the structure factors are then calculated in a straightforward way,
yielding:
\begin{equation}
\begin{aligned} &  & g_{s,h_{1}} & =0.602\ , & g_{s,h_{2}} & =\dfrac{2}{3}g_{h_{1}}\ , & g_{s,x} & =\dfrac{0.061-0.897\sin^{2}\phi_{x}}{\alpha^{2}+\sin^{2}\phi_{x}}\ , & g_{s,y} & =\dfrac{0.061-0.897\cos^{2}\phi_{y}}{\alpha^{2}+\cos^{2}\phi_{y}}\\
 &  & g_{d,h_{1}} & =-0.640\cos2\theta\ , & g_{d,h_{2}} & =-\dfrac{2}{3}g_{h_{1}}\ , & g_{d,x} & =-\dfrac{0.182+0.479\sin^{2}\phi_{x}}{\alpha^{2}+\sin^{2}\phi_{x}}\ , & g_{d,y} & =\dfrac{0.182+0.479\cos^{2}\phi_{y}}{\alpha^{2}+\cos^{2}\phi_{y}}
\end{aligned}
\label{Eqn:GapModel}
\end{equation}

Substitution in Eq. (\ref{gap_mixture}) and in the definition of
the anisotropic penetration depth (\ref{S_Delta_lambda}) gives the
result shown in Fig. 4 of the main text.

\section{Time-reversal symmetry breaking and orbital order}

In this section we discuss under what conditions the mixing parameter
$\gamma$ in Eq. (\ref{gap_mixture}) is real or complex. Note that
a comlex $\gamma$ effectively lifts the nodes that appear for a real
$\gamma$. To analytically address this question, we simplify our
three-band model further and ignore the angular dependence of the
pairing interaction. Thus, the effect of orbital order is included
only in the inter-pocket pairing interaction
\begin{equation}
\Gamma_{hx}=V(1+\varphi)\ ,\quad\Gamma_{hy}=V(1-\varphi)\ ,\mbox{ and}\quad\Gamma_{xy}=W\ ,\label{pairing}
\end{equation}
where $\varphi$, the nematic order parameter, is proportional to
the orbital order parameter (see also Ref. \cite{RMFSDMix}). For
concreteness, we set $\Delta_{h}$ to be real and positive, and write
$\Delta_{x}=\Delta_{1}\exp(i\theta_{1})$ and $\Delta_{y}=\Delta_{2}\exp(i\theta_{2})$,
with $\Delta_{1,2}\geq0$. In the absence of orbital order ($\varphi=0$),
the $s^{+-}$ state takes place for $V>W$ and is characterized by
$\theta_{1}=\theta_{2}=\pi$, whereas the $d$-wave state, taking
place for $V<W$, has $\theta_{1}=0$, $\theta_{2}=\pi$. Thus, any
solution in the presence of orbital order ($\varphi\neq0$) with $\theta_{i}\neq0,\pi$
necessarily implies that the mixing coefficient $\gamma$ in Eq. (\ref{gap_mixture})
is complex. Note that $\theta_{i}\neq0,\pi$ also implies that the
superconducting state is time-reversal symmetry-breaking (TRSB), since
$\Delta_{i}^{*}\neq\Delta_{i}$.

To proceed, we solve the $T=0$ BCS-like equations of this model.
By denoting $v=VN_{0}$ and $w=WN_{0}$, where $N_{0}$ is the density
of states at the Fermi energy, we have:
\begin{equation}
\left\{ \begin{aligned}\Delta_{h} & =-v(1+\varphi)\Delta_{1}\ln\frac{2\Lambda}{\Delta_{1}}\cos\theta_{1}-v(1-\varphi)\Delta_{2}\ln\frac{2\Lambda}{\Delta_{2}}\cos\theta_{2}\\
0 & =-v(1+\varphi)\Delta_{1}\ln\frac{2\Lambda}{\Delta_{1}}\sin\theta_{1}-v(1-\varphi)\Delta_{2}\ln\frac{2\Lambda}{\Delta_{2}}\sin\theta_{2}\\
\Delta_{1}\cos\theta_{1} & =-v(1+\varphi)\Delta_{h}\ln\frac{2\Lambda}{\Delta_{h}}-w\Delta_{2}\ln\frac{2\Lambda}{\Delta_{2}}\cos\theta_{2}\\
\Delta_{1}\sin\theta_{1} & =-w\Delta_{2}\ln\frac{2\Lambda}{\Delta_{2}}\sin\theta_{2}\\
\Delta_{2}\cos\theta_{2} & =-v(1-\varphi)\Delta_{h}\ln\frac{2\Lambda}{\Delta_{h}}-w\Delta_{1}\ln\frac{2\Lambda}{\Delta_{1}}\cos\theta_{1}\\
\Delta_{2}\sin\theta_{2} & =-w\Delta_{1}\ln\frac{2\Lambda}{\Delta_{1}}\sin\theta_{1}\ .
\end{aligned}
\right.\label{Eqn:TRSB_ThreeBand}
\end{equation}

Solving Eq.~\ref{Eqn:TRSB_ThreeBand}, we find that the TRSB solution
exists only when the following condition is satisfied
\begin{align}
 & \left\vert \left(1+\varphi\right)-\frac{w}{v}\exp\left(\frac{1+\varphi}{w(1-\varphi)}-\frac{w}{v^{2}(1-\varphi^{2})}\right)\right\vert <(1-\varphi)\exp\left(\frac{4\varphi}{w(1-\varphi^{2})}\right)<\nonumber \\
< & \ \left(1+\varphi\right)+\frac{w}{v}\exp\left(\frac{1+\varphi}{w(1-\varphi)}-\frac{w}{v^{2}(1-\varphi^{2})}\right)\label{Eqn::TRSBCondition}
\end{align}

In the tetragonal phase ($\varphi=0$), and in the weak-coupling limit
$w,v\ll1$, this condition reduces to
\begin{equation}
v-w<\frac{\ln2}{2}w^{2}\ ,\label{reduced_eq}
\end{equation}
in agreement with \cite{Valentin,Maiti}. It is instructive to compare
the TRS and TRSB solutions in this regime:
\begin{equation}
\mbox{TRS: }\left\{ \begin{aligned}\Delta_{h} & =0\\
\Delta_{x} & =2\Lambda\exp(-1/w)\\
\Delta_{y} & =-2\Lambda\exp(-1/w)
\end{aligned}
\right.\qquad\mbox{TRSB: }\left\{ \begin{aligned}\Delta_{h} & =2\Lambda\exp(-w/v^{2})\\
\left|\Delta_{x}\right| & =\left|\Delta_{y}\right|=2\Lambda\exp(-1/w)\\
\theta_{x} & =-\theta_{y}=\arccos\left(-\frac{w}{2v}\exp\left(\frac{1}{w}-\frac{w}{v^{2}}\right)\right)
\end{aligned}
\right.\ .\label{Eqn:TRSBNoOOSol}
\end{equation}

From Eq.~\ref{Eqn:TRSBNoOOSol}, it is clear that the TRSB state
has a larger condensation energy $\sum_{i}\left|\Delta_{i}\right|^{2}$
than the TRS state, and is therefore the global energy minimum. Note
however that in the regime $\frac{w}{v}-1\gg v$, $|\Delta_{h}|\ll|\Delta_{x/y}|$
and $\theta_{x}-\theta_{y}\approx\pi$. Therefore, as the system moves
farther from the $s^{+-}$/$d$-wave degeneracy point $w=v$, the
energy of the TRSB state asymptoticaly approaches the energy of the
TRS state.

\begin{figure}[htbp]
\begin{centering}
\includegraphics[height=2in]{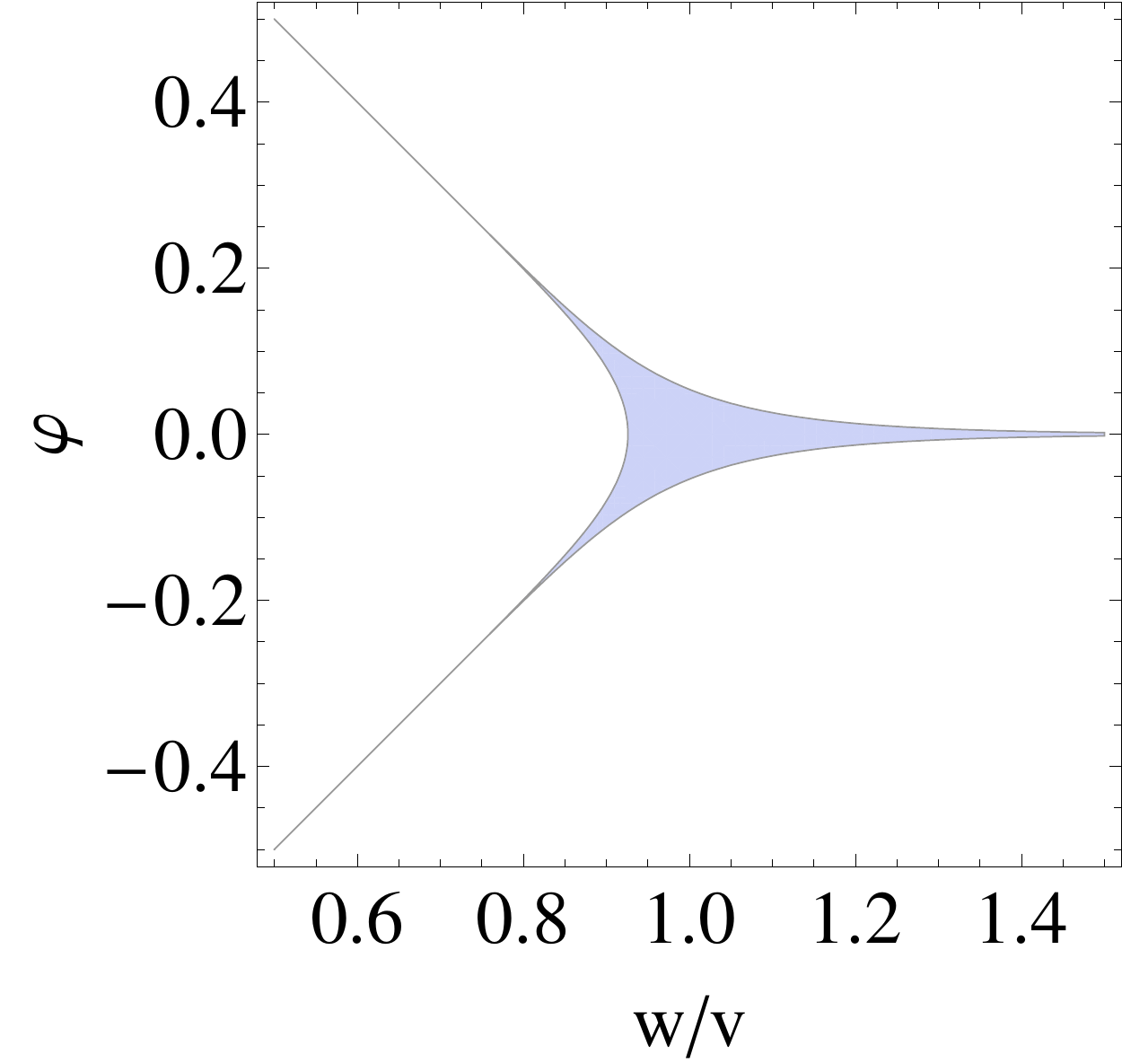} \caption{The shaded blue region, given by condition (\ref{Eqn::TRSBCondition}),
corresponds to the regime in which the superconducting ground state
breaks time-reversal symmetry. Here, $\varphi$ is proportional to
the orbital order parameter, and $w/v$ is the ratio between electron
pocket-electron pocket and hole pocket-electron pocket pairing interactions.
In this plot, $v=0.2$.}

\par\end{centering}

\label{Fig:TRSB}
\end{figure}

For non-zero orbital order, $\varphi\neq0$, we solve Eq. (\ref{Eqn::TRSBCondition})
numerically to find the region in the $\left(\frac{w}{v},\varphi\right)$
phase diagram for which the superconducting state is TRSB. As shown
in Fig.~\ref{Fig:TRSB}, we find that orbital order in general suppresses
the TRSB phase, which is restricted to the vicinity of the degeneracy
point $w=v$. Note that, along the three very thin long branches in
the plot, the condensation energy of the TRSB solution is lower but
very close to the energy of the TRS solution (with $\theta_{1,2}=0,\pi$).
The very long extension of these branches is an artifact of our simplified
model, which can be eliminated if one includes angular-dependent terms
in the pairing interaction (\ref{pairing}).


\begin{thebibliography}{10}
\bibitem{LaOFeAs} Y. Kamihara, T. Watanabe, M. Hirano, and H. Hosono,
J.\ Am.\ Chem.\ Soc. \textbf{130}, 3296 (2008).

\bibitem{BaFe2As2} M. Rotter, M. Tegel, D. Johrendt, Phys. Rev. Lett.
\textbf{101}, 107006 (2008).

\bibitem{DingARPES122K} H. Ding, P. Richard, K. Nakayama, K. Sugawara,
T. Arakane, Y. Sekiba, A. Takayama, S. Souma, T. Sato, T. Takahashi,
Z. Wang, X. Dai, Z. Fang, G. F. Chen, J. L. Luo, and N. L. Wang, Europhys.
Lett. \textbf{83}, 47001 (2008).

\bibitem{Shimojima_Kdoped} T. Shimojima, F. Sakaguchi, K. Ishizaka,
Y. Ishida, T. Kiss, M. Okawa, T. Togashi, C.T. Chen, S. Watanabe,
M. Arita, K. Shimada, H. Namatame, M. Taniguchi, K. Ohgushi, S. Kasahara,
T. Terashima, T. Shibauchi, Y. Matsuda, A. Chainani, and S. Shin,
Science \textbf{332}, 564 (2011).

\bibitem{Prozorov122CoGap} M. A. Tanatar, J.-Ph. Reid, H. Shakeripour,
X. G. Luo, N. Doiron-Leyraud, N. Ni, S. L. Bud'ko, P. C. Canfield,
R. Prozorov, and L. Taillefer, Phys. Rev. Lett. \textbf{104}, 067002
(2010).

\bibitem{BorisenkoLi111Gap} S.V. Borisenko, V. B. Zabolotnyy, A.A.
Kordyuk, D.V. Evtushinsky, T.K. Kim, I.V. Morozov, R. Follath, and
B. Büchner, Symmetry \textbf{4}, 251 (2012).

\bibitem{AllanLiFeAs} M. P. Allan, A. W. Rost, A. P. Mackenzie, Yang
Xie, J. C. Davis, K. Kihou, C. H. Lee, A. Iyo, H. Eisaki, T.-M. Chuang,
Science \textbf{336}, 563 (2012).

\bibitem{Matsuda122PNode} K. Hashimoto, M. Yamashita, S. Kasahara,
Y. Senshu, N. Nakata, S. Tonegawa, K. Ikada, A. Serafin, A. Carrington,
T. Terashima, H. Ikeda, T. Shibauchi, and Y. Matsuda, Phys. Rev. B
\textbf{81}, 220501(R) (2010).

\bibitem{FengPnode} Y. Zhang, Z. R. Ye, Q. Q. Ge, F. Chen, Juan Jiang,
M. Xu, B. P. Xie, and D. L. Feng, Nature Phys. \textbf{8}, 371 (2012).

\bibitem{Pdoped_nodes} T. Yoshida, S. Ideta, T. Shimojima, W. Malaeb,
K. Shinada, H. Suzuki, I. Nishi, A. Fujimori, K. Ishizaka, S. Shin,
Y. Nakashima, H. Anzai, M. Arita, A. Ino, H. Namatame, M. Taniguchi,
H. Kumigashira, K. Ono, S. Kasahara, T. Shibauchi, T. Terashima, Y.
Matsuda, M. Nakajima, S. Uchida, Y. Tomioka, T. Ito, K. Kihou, C.
H. Lee, A. Iyo, H. Eisaki, H. Ikeda, R. Arita, T. Saito, S. Onari,
and H. Kontani, arXiv:1301.4818.

\bibitem{Ru_doped_nodes} X. Qiu, S. Y. Zhou, H. Zhang, B. Y. Pan,
X. C. Hong, Y. F. Dai, Man Jin Eom, Jun Sung Kim, Z. R. Ye, Y. Zhang,
D. L. Feng, and S. Y. Li, Phys. Rev. X \textbf{2}, 011010 (2012).

\bibitem{Xue11Node} C.-L. Song, Y.-L. Wang, Y.-P. Jiang, W. Li, T.
Zhang, Z. Li, K. He, L. Wang, J.-F. Jia, H.-H. Hung, C. Wu, X. Ma,
X. Chen, and Q.-K. Xue, Science \textbf{332}, 1410 (2011).

\bibitem{MatsudaK122Node} K. Hashimoto, A. Serafin, S. Tonegawa,
R. Katsumata, R. Okazaki, T. Saito, H. Fukazawa, Y. Kohori, K. Kihou,
C. H. Lee, A. Iyo, H. Eisaki, H. Ikeda, Y. Matsuda, A. Carrington,
and T. Shibauchi, Phys. Rev. B \textbf{82}, 014526 (2010).

\bibitem{ShinK122Node} K. Okazaki, Y. Ota, Y. Kotani, W. Malaeb,
Y. Ishida, T. Shimojima, T. Kiss, S. Watanabe, C.-T. Chen, K. Kihou,
C. H. Lee, A. Iyo, H. Eisaki, T. Saito, H. Fukazawa, Y. Kohori, K.
Hashimoto, T. Shibauchi, Y. Matsuda, H. Ikeda, H. Miyahara, R. Arita,
A. Chainani, and S. Shin, Science \textbf{337}, 1314 (2012).

\bibitem{TailleferK122Pairing} F. F. Tafti, A. Juneau-Fecteau, M.-E.
Delage, S. Rene de Cotret, J.-Ph. Reid, A. F. Wang, X.-G. Luo, X.
H. Chen, N. Doiron-Leyraud, and L. Taillefer, Nature Physics \textbf{9},
349 (2013).

\bibitem{Johrendta122CoP} V. Zinth and D. Johrendt, Europhys. Lett.
\textbf{98}, 57010 (2012).

\bibitem{magnetic} I. I. Mazin, D. J. Singh, M. D. Johannes, and
M. H. Du, Phys. Rev. Lett. \textbf{101}, 057003 (2008); A. V. Chubukov,
D. V. Efremov and I Eremin, Phys. Rev. B \textbf{78}, 134512 (2008);
K. Kuroki, S. Onari, R. Arita, H. Usui, Y. Tanaka, H. Kontani, and
H. Aoki, Phys. Rev. Lett. \textbf{101}, 087004 (2008); V. Cvetkovi\'{c}
and Z. Tešanovi\'{c}, Phys. Rev. B \textbf{80}, 024512 (2009); J.
Zhang, R. Sknepnek, R. M. Fernandes, and J. Schmalian, Phys. Rev.
B \textbf{79}, 220502(R) (2009).

\bibitem{reviews_pairing} A. V. Chubukov, Annu. Rev. Cond. Mat. Phys.
\textbf{3}, 57 (2012); P. J. Hirschfeld, M. M. Korshunov, and I. I.
Mazin, Rep. Prog. Phys. \textbf{74}, 124508 (2011).

\bibitem{DingARPESOrbital} X.-P. Wang, P. Richard, Y.-B. Huang, H.
Miao, L. Cevey, N. Xu, Y.-J. Sun, T. Qian, Y.-M. Xu, M. Shi, J.-P.
Hu, X. Dai, and H. Ding, Phys. Rev. B \textbf{85}, 214518 (2012).

\bibitem{Shen122CoOO} M. Yi, D. Lu, J.-H. Chu, J. G. Analytis, A.
P. Sorini, A. F. Kemper, B. Moritz, S.-K. Mo, R. G. Moore, M. Hashimoto,
W.-S. Lee, Z. Hussain, T. P. Devereaux, I. R. Fisher, and Z.-X. Shen,
PNAS, \textbf{108}, 6878 (2011).

\bibitem{Kuroki09} K. Kuroki, H. Usui, S. Onari, R. Arita, and H.
Aoki, Phys. Rev. B \textbf{79}, 224511 (2009).

\bibitem{GraserSDDeg} S. Graser, T. A. Maier, P. J. Hirschfeld, and
D. J. Scalapino, New J. Phys. \textbf{11}, 025016 (2009).

\bibitem{Chubukov09} A. V. Chubukov, M. G. Vavilov, and A. B. Vorontsov,
Phys. Rev. B \textbf{80}, 140515(R) (2009).

\bibitem{CWu09} W.-C. Lee, S.-C. Zhang, and C. Wu, Phys. Rev. Lett.
\textbf{102}, 217002 (2009).

\bibitem{KemperNJP} A. F. Kemper, T. A. Maier, S. Graser, H.-P. Cheng,
P. J. Hirschfeld, D. J. Scalapino, New J. Phys. \textbf{12}, 073030
(2010).

\bibitem{Graser10} S. Graser, A. F. Kemper, T. A. Maier, H.-P. Cheng,
P. J. Hirschfeld, and D. J. Scalapino, Phys. Rev. B \textbf{81}, 214503
(2010).

\bibitem{Ikeda10} H. Ikeda, R. Arita, and J. Kunes, Phys. Rev. B
\textbf{81}, 054502 (2010).

\bibitem{Wang10} F. Wang, H. Zhai, and D.-H. Lee, Phys. Rev. B \textbf{81},
184512 (2010).

\bibitem{Maiti11} S. Maiti, M. M. Korshunov, T. A. Maier, P. J. Hirschfeld,
and A. V. Chubukov, Phys. Rev. B \textbf{84}, 224505 (2011); \emph{ibid}
Phys. Rev. Lett. \textbf{107}, 147002 (2011).

\bibitem{Thomale11} R. Thomale, C. Platt, W. Hanke, J. Hu, and B.
A. Bernevig, Phys. Rev. Lett. \textbf{107}, 117001 (2011).

\bibitem{Fernandes13} R. M. Fernandes and A. J. Millis, Phys. Rev.
Lett. \textbf{110}, 117004 (2013).

\bibitem{Kotliar13} Z. P. Yin, K. Haule, and G. Kotliar, arxiv:1311.1188.

\bibitem{raman_mode1} F. Kretzschmar, B. Muschler, T. Böhm, A. Baum,
R. Hackl, H.-H. Wen, V. Tsurkan, J. Deisenhofer, and A. Loidl, Phys.
Rev. Lett. \textbf{110}, 187002 (2013).

\bibitem{raman_mode2} M. Khodas, A. V. Chubukov, and G. Blumberg,
arXiv:1405.6246.

\bibitem{FisherXnemDivergent} J.-H. Chu, H.-H. Kuo, J. G. Analytis,
and I. R. Fisher, Science \textbf{337}, 710 (2012).

\bibitem{Kuo13} H.-H. Kuo, M. C. Shapiro, S. C. Riggs, and I. R.
Fisher, Phys. Rev. B \textbf{88}, 085113 (2013).

\bibitem{shear_modulus} R. M. Fernandes, L. H. VanBebber, S. Bhattacharya,
P. Chandra, V. Keppens, D. Mandrus, M. A. McGuire, B. C. Sales, A.
S. Sefat, and J. Schmalian, Phys. Rev. Lett. \textbf{105}, 157003
(2010).

\bibitem{Yoshizawa12} M. Yoshizawa, D. Kimura, T. Chiba, A. Ismayil,
Y. Nakanishi, K. Kihou, C.-H. Lee, A. Iyo, H. Eisaki, M. Nakajima,
and S. Uchida, J. Phys. Soc. Jpn. \textbf{81}, 024604 (2012).

\bibitem{Matsuda12} S. Kasahara, H. J. Shi, K. Hashimoto, S. Tonegawa,
Y. Mizukami, T. Shibauchi, K. Sugimoto, T. Fukuda, T. Terashima, A.
H. Nevidomskyy, and Y. Matsuda, Nature \textbf{486}, 382 (2012).

\bibitem{Meingast122Xnem} A. E. Böhmer, P. Burger, F. Hardy, T. Wolf,
P. Schweiss, R. Fromknecht, M. Reinecker, W. Schranz, and C. Meingast,
Phys. Rev. Lett. \textbf{112}, 047001 (2014).

\bibitem{GallaisRamanNem} Y. Gallais, R. M. Fernandes, I. Paul, L.
Chauviere, Y.-X. Yang, M.-A. Measson, M. Cazayous, A. Sacuto, D. Colson,
and A. Forget, Phys. Rev. Lett. \textbf{111}, 267001 (2013).

\bibitem{Jigang14} A. Patz \emph{et al}., Nature Comm. \textbf{5},
3229 (2014).

\bibitem{PasupathyNa111NemAFM} E. P. Rosenthal, E. F. Andrade, C.
J. Arguello, R. M. Fernandes, L. Y. Xing, X. C. Wang, C. Q. Jin, A.
J. Millis, and A. N. Pasupathy, Nature Phys. \textbf{10}, 225 (2014).

\bibitem{Fisher10} J.-H. Chu, J. G. Analytis, K. De Greve, P. L.
McMahon, Z. Islam, Y. Yamamoto, and I. R. Fisher, Science \textbf{329},
824 (2010).

\bibitem{Tanatar10} M. A. Tanatar, E. C. Blomberg, A. Kreyssig, M.
G. Kim, N. Ni, A. Thaler, S. L. Bud'ko, P. C. Canfield, A. I. Goldman,
I. I. Mazin, and R. Prozorov, Phys. Rev. B \textbf{81}, 184508 (2010).

\bibitem{Degiorgi12} A. Lucarelli, A. Dusza, A. Sanna, S. Massidda,
J.-H. Chu, I.R. Fisher, and L. Degiorgi, New J. Phys. \textbf{14},
023020 (2012).

\bibitem{Dhital12} C. Dhital \emph{et al.,} Phys. Rev. Lett. \textbf{108},
087001 (2012).

\bibitem{RMFRevNem} R. M. Fernandes, A. V. Chubukov, and J. Schmalian,
Nature Phys. \textbf{10}, 97 (2014).

\bibitem{RMFSDMix} R. M. Fernandes, and A. J. Millis, Phys. Rev.
Lett. \textbf{111}, 127001 (2013); F. Yang, F. Wang, and D.-H. Lee,
Phys. Rev. B \textbf{88}, 100504 (2013).

\bibitem{Yi12} M. Yi \emph{et al}., New J. Phys. \textbf{14}, 073019
(2012).

\bibitem{Zhang12} Y. Zhang \emph{et al.}, Phys Rev B \textbf{85},
085121 (2012).

\bibitem{w_ku10} C. C. Lee, W. G. Yin, and W. Ku, Phys. Rev. Lett.
\textbf{103}, 267001 (2009).

\bibitem{Devereaux10} C.-C. Chen, J. Maciejko, A. P. Sorini, B. Moritz,
R. R. P. Singh, and T. P. Devereaux, Phys. Rev. B \textbf{82}, 100504
(2010).

\bibitem{Phillips12} W.-C. Lee and P. W. Phillips, Phys. Rev. B \textbf{86},
245113 (2012).

\bibitem{Kontani12} S. Onari H. and Kontani, Phys. Rev. Lett. \textbf{109},
137001 (2012).

\bibitem{Dagotto13} S. Liang, A. Moreo, and E. Dagotto, Phys. Rev.
Lett. \textbf{111}, 047004 (2013).

\bibitem{Fernandes12} R. M. Fernandes, A. V. Chubukov, J. Knolle,
I. Eremin, and J. Schmalian, Phys. Rev. B \textbf{85}, 024534 (2012).

\bibitem{Lv11} W. Lv and P. Phillips, Phys. Rev. B \textbf{84}, 174512
(2011).

\bibitem{Vafek13} V. Cvetkovic and O. Vafek, Phys. Rev. B \textbf{88},
134510 (2013)

\bibitem{Valentin} V. Stanev and Z. Tesanovic, Phys. Rev. B \textbf{81},
134522 (2010).

\bibitem{Maiti} S. Maiti and A. V. Chubukov, Phys. Rev. B \textbf{87},
144511 (2013).

\bibitem{quadratic_nodes1} R. M. Fernandes and J. Schmalian, Phys.
Rev. B \textbf{84}, 012505 (2011).

\bibitem{quadratic_nodes2} V. Stanev, B. S. Alexandrov, P. Nikolic,
and Z. Tesanovic, Phys. Rev. B \textbf{84}, 014505 (2011).

\bibitem{quadratic_nodes3} B. Mazidian, J. Quintanilla, A. D. Hillier,
and J. F. Annett, Phys. Rev. B \textbf{88}, 224504 (2013).

\bibitem{Matsuda_penetration_depth} K. Hashimoto, K. Cho, T. Shibauchi,
S. Kasahara, Y. Mizukami, R. Katsumata, Y. Tsuruhara, T. Terashima,
H. Ikeda, M. A. Tanatar, H. Kitano, N. Salovich, R. W. Giannetta,
P. Walmsley, A. Carrington, R. Prozorov, and Y. Matsuda, Science \textbf{336},
1554 (2012).

\bibitem{Kasahara10} S. Kasahara \emph{et al.}, Phys. Rev. B \textbf{81},
184519 (2010).
\end{thebibliography}
\end{document}